\newcommand{\nh}{N$_{\rm H}$}

\newcommand{\ergs}{erg\,s$^{-1}$}
\newcommand{\kms}{km\,s$^{-1}$}
\newcommand{\masyr}{mas\,yr$^{-1}$}

\newcommand{\Lx}{$\rm L_{\rm X}$}

\newcommand{\rxdixhuit}{\object{RX\,J1856.5\-$-$3754}}
\newcommand{\rxzerosept}{\object{RX\,J0720.4\-$-$3125}}
\newcommand{\rxseize}{\object{RX\,\-J16\-05.3\-+3249}}
\newcommand{\rbsdouze}{\object{RX\,J1308.6\-+2127}}
\newcommand{\rbsdixsept}{\object{RX\,J2143.0\-+0654}}
\newcommand{\rxzerohuit}{\object{RX\,J0806.4\-$-$4123}}
\newcommand{\rxzeroquatre}{\object{RX\,\-J0420.0\-$-$5022}}
\newcommand{\degr}{\ensuremath{^\circ}}

\documentclass[traditabstract]{aa} 
\usepackage{txfonts}
\usepackage{epsfig}
\usepackage{rotating}
\usepackage{natbib}
\begin{document}
   \title{Proper motions of thermally emitting isolated neutron stars measured with Chandra}

   \subtitle{}
   
   \author{
   C. Motch \inst{1}    
   \and
   A. M. Pires \inst{1,2}
   \and
   F. Haberl  \inst{3}
   \and
   A. Schwope \inst{4}
   \and
   V.E. Zavlin \inst{5}
   }
    
   \institute{
   CNRS, Observatoire Astronomique, Universit\'e de Strasbourg, 11 rue de l'Universit\'e, F-67000, Strasbourg, France \\
   \email{motch@astro.u-strasbg.fr}
   \and
   Instituto de Astronomia, Geof\'isica e Ci\^encias Atmosf\'ericas, Universidade de S\~ao Paulo, R. do Mat\~ao 1226, 05508-090 S\~ao Paulo, Brazil
   \and
   Max-Planck-Institut f\"ur extraterrestrische Physik, Giessenbachstrasse, D-85748 Garching, Germany
   \and
   Astrophysikalisches Institut Potsdam, An der Sternwarte 16, D-14482 Potsdam, Germany 
   \and 
   Space Science Laboratory, Universities Space Research Association, NASA MSFC VP62, Huntsville, AL 35805, USA
   }
   
   \date{Received ; accepted }

\abstract{The remarkable astrometric capabilities of the Chandra Observatory offer the possibility to measure proper motions of X-ray sources with an unprecedented accuracy in this wavelength range. We recently completed a proper motion survey of three of the seven thermally emitting radio-quiet isolated neutron stars (INSs) discovered in the ROSAT all-sky survey. These INSs (\rxzeroquatre, \rxzerohuit\ and \rbsdouze ) either lack an optical counterpart or have one so faint that ground based or space born optical observations push the current possibilities of the instrumentation to the limit . Pairs of ACIS observations were acquired 3 to 5 years apart to measure the displacement of the sources on the X-ray sky using as reference the background of extragalactic or remote Galactic X-ray sources. We derive 2\,$\sigma$ upper limits of 123\,\masyr\ and 86\,\masyr\ on the proper motion of \rxzeroquatre\ and \rxzerohuit, respectively. \rbsdouze\ exhibits a very significant displacement ($\sim$ 9 $\sigma$) yielding $\mu$ = 220\,$\pm$\,25\,\masyr, the second fastest measured among all ROSAT discovered INSs. The source is probably moving away rapidly from the Galactic plane at a speed which precludes any significant accretion of matter from the interstellar medium. Its transverse velocity of $\sim$ 740 ($d$/700\,pc)\,km/s might be the largest of all ROSAT INSs and its corresponding spatial velocity stands among the fastest recorded for neutron stars. \rbsdouze\ is thus a  middle-aged (age $\sim$ 1\,My) high velocity cooling neutron star. We investigate its possible origin in nearby OB associations or from a field OB star. In most cases, the flight time from birth place appears significantly shorter than the characteristic age derived from spin down rate. Overall, the distribution in transverse velocity of the ROSAT INSs is not statistically different from that of normal radio pulsars.
}

               
\maketitle

%

\section{Introduction}

Observational studies of neutron stars started more than forty years ago with the discovery of radio pulsars. Owing to the large luminosity released by accretion, neutron stars in binaries were among the first sources detected by the pioneering X-ray instruments. The advent of sensitive imaging X-ray telescopes such as Einstein, ROSAT and now Chandra and XMM-Newton has revealed the complex pattern of high energy emission from isolated neutron stars (INSs). More than forty radio pulsars have been detected in X-rays, with an X-ray luminosity ranging from $10^{-5}$ to $10^{-1}$ of their spin-down luminosity \citep{kargaltsev2008}. Although radio surveys are still the main providers of new INSs, X-ray observations are continuously revealing the existence of neutron stars undetected in the radio regime. These radio-quiet neutron stars are unlikely to constitute an homogeneous class. Among these are central compact objects (CCO) in SNRs. These soft X-ray sources are characterized by a steady thermal X-ray emission with a temperature in the range of 0.2 to 0.5\,keV (see \cite{pavlov2004} and \cite{deluca2008a} for recent reviews). The ages of their associated SNR are of a few kyr. Two CCOs exhibit pulsations periods of a fraction of a second and their period derivative points at low surface magnetic fields of less than a few 10$^{11}$\,G \citep{gotthelf2008}. Anomalous X-ray pulsars (AXPs) and soft-$\gamma$ repeaters (SGRs) form other classes of INSs revealed through their luminous X-ray emission. While SGRs appear as young as CCOs, their strong magnetic fields (B $\ga$ 10$^{14}$G) drives a wealth of outstanding magnetospheric processes (see \cite{wood2006} for a recent review). ROSAT was instrumental in discovering the X-ray thermal emission of seven isolated neutron stars. Nicknamed "The Magnificent Seven", X-ray dim or X-ray thermally emitting isolated neutron stars (depending on authors' preferences), we will simply refer to them as ROSAT discovered INSs. These INSs exhibit overall properties that are at variance with those of most radio pulsars. Deep searches failed to detect any radio emission \citep{kondra2008} although unconfirmed claims of detection at long wavelengths exist \citep{malo2007}. Their X-ray spectra are essentially thermal with $kT$ in the range of 40 to 100\,eV. Most ROSAT INSs also display shallow broad absorption features at energies around 300\,eV which can be interpreted as proton cyclotron lines or as atomic H or He transitions in high magnetic field conditions ($\geq$ few 10$^{13}$ G) \citep{ho2004}. High proper motions in excess of 100\,mas\,yr$^{-1}$ rule out accretion from the interstellar medium as a significant powering mechanism and thus leave no doubt that these INSs are young or middle-aged (age $\lesssim$ 10$^{6}$\,yr) cooling objects. Their X-ray spectra undergo little photoelectric absorption indicating distances of a few hundred parsec, confirmed in two cases by HST parallaxes. With the exception of \rxseize\ for which no periodicity is detected yet, all exhibit rather long spin periods in the range of 3.4\,s to 11.4\,s. When measured, the magnetic field implied by the spin down rate is roughly consistent with that inferred from the low energy absorption lines (see \cite{haberl2007} and \cite{kaplan2008} for recent reviews).

Young neutron stars are expected to be strong X-ray sources, the high energy emission being either due to cooling or to non-thermal mechanisms. X-rays are thus an unbiased mean to detect nearby neutron stars. In this respect, it is striking that ROSAT discovered INSs appear to represent about half of all young (age $\leq 3 \times 10^{6}$\,yr) INSs known within $\sim$ 1\,kpc \citep{popov2003}. ROSAT discovered INSs could thus be the tip of the iceberg of a formerly hidden large population of stellar remnants which may be, at least locally, as numerous as radio pulsars. 

Their properties lie at the crossroads between those of other groups of INSs. High magnetic fields and possible evidences for significant magnetospheric emission detected in the optical/UV bands may establish an evolutionary link between these objects and the group of magnetars (AXPs and SGRs) \citep{zane2008}. Some long period and high B radio pulsars share the same locus as ROSAT discovered INSs in the $P$/$\dot{P}$ diagram, suggesting that the lack of radio emission could be due to the fact that the radio beam which narrows at long spin period does not sweep over the earth. Alternatively, the mechanism quenching radio emission could be similar to that transitively at work in rotating radio transients (RRATs) \citep{mclaughlin2006}. 

Proper motion measurements provide a number of important diagnostics on population properties. First, the relative velocities with respect to the interstellar medium put constraints on the contribution of accretion to the overall energy budget. Second, the reconstruction of backward trajectories allows to estimate the dynamical age of the neutron star assuming a birth in the Galactic plane and more precisely in one of the nearby OB associations. The flight time can then be compared with the age derived from spin down, revealing a possible non standard braking mechanism or pointing toward a long spin period at birth. Finally, the dynamical age also constrains the cooling mechanism \citep{yakovlev2004}.

Proper motions in the range of 108 to 333\,mas\,yr$^{-1}$ have been measured for the three X-ray brightest ROSAT discovered INSs which have bright enough optical counterparts with $m_B$ $\lesssim$ 27 to allow measurements with large telescopes on the ground or with the HST \citep{kaplan2002b,motch2003,motch2005,zane2006,kaplan2007}. In addition, the high astrometric accuracy of the HST also allows to derive parallaxes for the nearest members \citep{kaplan2002b,kaplan2007}. However the remaining ROSAT INSs either have no known optical counterpart or have an optical candidate too faint ($m_B$ $\geq$ 28) to allow repeated observations in a reasonable amount of time. 
X-ray imaging telescopes of the former generation had an insufficient spatial resolution to detect source displacements on the sky in a reasonable interval of time, even for high proper motion objects such as INSs. The advent of high spatial resolution X-ray observatories has made possible the measurement with attracting accuracy of the proper motions of objects too faint to be detected at other wavelengths. Radio-quiet neutron stars are obviously prime targets for this kind of studies. 
We thus took advantage of the unprecedented spatial resolution of the Chandra Observatory to start in 2002 first epoch measurements of three of the X-ray faintest members, \rxzerohuit, \rxzeroquatre\ and \rbsdouze\, with second epoch taking place three to five years later. Observing close to the maximum of the spectral energy distribution Chandra provides a fast and efficient way to measure accurate relative positions with respect to the background of active galactic nuclei. In this paper we report on the results of this observing campaign, extensively discuss the error budget on the proper motion and present the implications of our findings on our understanding of the properties of this class of neutron stars. Preliminary results were presented in \cite{motch2007} and \cite{motch2008}.
 
\section{Target properties}

\subsection{\rxzerohuit}

The source was discovered by \cite{haberl1998} in the ROSAT all-sky survey. XMM-Newton observations revealed pulsations at a period of 11.4\,s \citep{haberl2002,haberl2004}, the longest recorded so far for this group of objects. With a best fit blackbody temperature of $kT$ = 96\,eV, \rxzerohuit\ lies among the hottest ROSAT discovered INSs. The spectral fit is significantly improved by adding a shallow absorption line centered around 460\,eV. The interstellar absorption toward the source is \nh\ = 4$\times$10$^{19}$\,cm$^{-2}$ or 1.1$\times$10$^{20}$\,cm$^{-2}$ depending on whether the low-energy broad absorption line is included or not in the spectral fit. \rxzerohuit\ is the ROSAT INS located at the lowest Galactic latitude ($b$ = -4.98\degr). Its modest photoelectric absorption suggests small distances of the order of 240\,pc \citep{posselt2007}. No optical counterpart brighter than a non constraining limit of $m_B$ $\sim$ 24 is present in the error circle \citep{haberl1998}. 

\subsection{\rxzeroquatre}

\rxzeroquatre\ is the X-ray faintest INS discovered in the ROSAT all-sky survey \citep{haberl1999}. With $kT$ = 45\,eV,  \rxzeroquatre\ is the coolest of its group. It also displays a broad absorption line at E $\sim$ 330\,eV \citep{haberl2004}. The source is located at relatively high galactic latitude ($b$ = -44.39\degr) and the interstellar absorption toward \rxzeroquatre\  is \nh\ = 1.0$\times$10$^{20}$\,cm$^{-2}$ or 2.0$\times$10$^{20}$\,cm$^{-2}$, again depending on the inclusion or not of a broad line in the spectral model, and is on average about twice that of \rxzerohuit. XMM-Newton EPIC observations have revealed a pulsation period of 3.45\,s, the shortest of all ROSAT discovered INSs. \cite{haberl2004} also reported the possible existence of a $m_B$ = 26.6$\pm$0.3 mag object in the Chandra error circle. However, the presence of a relatively bright star close to the INSs adds a lot of scattered optical light at the source position and makes deep imaging difficult. The distance estimate by \cite{posselt2007} on the basis of interstellar absorption is of the order of 345\,pc. 

\subsection{\rbsdouze}

\rbsdouze\ (alias RBS\,1223) is the fifth X-ray brightest of the INSs discovered by ROSAT. It exhibits one of the highest temperatures of the group $kT$ $\sim$ 102\,eV \citep{schwope2007} and a pulse period of 10.31\,s. A pair of possible harmonic proton cyclotron absorption lines are detected at $\sim$ 0.23 and $\sim$ 0.46\,keV which would indicate a magnetic field of the order of 4$\times$10$^{13}$\,G \citep{schwope2007}. However, the ratio of the line fluxes does not match that expected from cyclotron absorption and alternative interpretations in terms of two distinct poles or atomic transitions may also be considered. \cite{kaplan2005} could connect pulse phases observed by Chandra and XMM-Newton over a five year long time interval and derive a spin down rate of $\dot{P}$ = 1.12$\times$10$^{-13}$\,s/s. If due to magnetic dipole radiation $\dot{P}$ implies B = 3.4$\times$10$^{13}$\,G, a value roughly consistent with that estimated from the low energy absorption lines and the spin down age is $\sim$ 1.5$\times$10$^{6}$\,yr. The optical counterpart of \rbsdouze\ is likely a faint $m_V$ $\sim$ 28.6 object \cite{kaplan2002a} detected thanks to deep HST observations. The source is located at high galactic latitude ($b$ = +83.08\degr).

\section{Observations}

Up to now, only few attempts have been made to measure proper motions in X-rays. The first example is the confirmation of the high proper motion of \rxdixhuit\ (0.34 $\pm$ 0.12 \masyr, \cite{neu2001}) based on two ROSAT HRI observations obtained three years apart. A second case is the derivation of an upper limit of $\sim$ 170\,\masyr\ on the motion of AXP 1E2259+586 \citep{ot2005} using ROSAT, Chandra and XMM-Newton data. Another study bears on the central compact object in Puppis-A for which \cite{hui2006} derive a proper motion of 107.5 $\pm$ 34.5\,\masyr\ by comparing two Chandra HRC-I observations. Very recently, \cite{deluca2008b} and \cite{kaplan2008b} using Chandra ACIS-I observations reported 90\% upper limits on the proper motion of two magnetars, SGR 1900+14 and 1E 2259+586, of 54\,\masyr\ and 65\,\masyr, respectively. 
  
Our observing campaign started in Chandra Cycle 3 (2002) during which we obtained first epoch images for the three targets, \rxzerohuit, \rxzeroquatre\ and \rbsdouze. Second epoch observations took place in 2005 and 2007 (see Journal of Observations in Table \ref{jo}). We selected the ACIS cameras because of their superior hard X-ray sensitivities compared to that of HRC-I. We also thought that the impossibility to filter X-ray events according to energy would possibly degrade the centering accuracy of X-ray sources with HRC-I. Finally, at the time we had to take the decision, the fact that HRC pixels are determined by the electronic readout was felt as being a risk for the long term stability of the astrometry. Our goal was to use the background of extragalactic sources, mainly AGN, to define an astrometric reference frame relative to which we could measure the displacement of the isolated neutron star. In the case of the low Galactic latitude field of \rxzerohuit\ ($b\,\sim\,$--4.98\degr) some Galactic active stars are also expected to contribute to the reference frame. However, being active stars (mostly younger than 1-2\,Gyr ; \cite{motch1997}), their proper motion is small. Stars younger than 1-2\,Gyr still share most of the motion of their parent molecular clouds and thus have a small random velocity dispersion of less than $\sim$ 20\,\kms \citep{wielen1977}. Being low luminosity X-ray sources (\Lx $\la$ 10$^{31}$\ergs), the active coronae detected by Chandra are located at relatively short distances ($\sim$ 1\,kpc). Accordingly, the expected proper motion resulting from velocity dispersion and Galactic differential rotation is of the order of 22\,\masyr\ at 300\,pc for \rxzerohuit. Averaged over the whole ACIS-I field of view, the effect of young star displacements on relative astrometry is therefore negligible compared to the overall error budget.  

First {\em MARX}\footnote{http://space.mit.edu/CXC/MARX/} simulations carried out in 2001 indicated that an exposure time of 20\,ks was sufficient to detect a suitable number ($\geq$ 10) of well localized background reference sources and obtain an astrometric error of $\sim$ 0.07\arcsec\ on the relative positioning of the two epochs. Because of the enhanced soft X-ray sensitivity of ACIS-S compared to ACIS-I, the two X-ray brightest targets, \rxzerohuit\ and \rbsdouze\ would have suffered from pile-up in ACIS-S and would thus have likely produced biased X-ray positions. Using ACIS-I for these two sources provided the best compromise between the need to observe enough background sources and the requirement to have the best position for the INS target. However, for the fainter and much softer X-ray source \rxzeroquatre , ACIS-I would have collected too few counts.  Accordingly we chose to use ACIS-S at the expense of a smaller field of view. The total source count rates were of 6$\times$10$^{-2}$, 6$\times$10$^{-2}$ and 1.1$\times$10$^{-1}$ cts\,s$^{-1}$ for \rxzerohuit, \rxzeroquatre\ and \rbsdouze\ respectively implying a pile-up fraction of less then 3\% in the worst case.

Our observing strategy was to enforce the same acquisition mode, roll angle and aimpoints at both epochs so as to minimize the impact of a different orientation on the sky or camera settings on the final accuracy of the relative astrometry.  

\begin{table}[t]
\caption{Journal of Observations}
\centering
\label{obs}       
\begin{tabular}{lcccc}
\hline\noalign{\smallskip}
Object & Observation & Camera & Exposure & Roll angle\\[3pt]
       & date        &        & time (ks)& (deg)\\[3pt]
\rxzerohuit   & 2002-02-23 & ACIS-I & 17.7 & 322.4  \\
\rxzerohuit   & 2005-02-18 & ACIS-I & 19.7 & 325.6  \\
\rxzeroquatre & 2002-11-11 & ACIS-S & 19.4 & 18.6 \\
\rxzeroquatre & 2005-11-07 & ACIS-S & 19.7 & 23.6 \\
\rbsdouze     & 2002-05-21 & ACIS-I & 19.5 & 226.5 \\
\rbsdouze     & 2007-05-12 & ACIS-I & 20.2 & 226.7 \\

\noalign{\smallskip}\hline
\end{tabular}
\label{jo}
\end{table}

\section{Data analysis}

\subsection{Source detection}

Data reduction was performed in two steps. The 2002 and 2005 observations of \rxzerohuit\ and \rxzeroquatre\ were analysed using CIAO~3.3.0.1 while the 2002 and 2007 data of \rbsdouze\ were reduced with CIAO~3.4 and CIAO~4.0. In all cases we used the most recently available calibration database. We corrected remaining attitude errors using the aspect calculator available in the science threads\footnote{http://asc.harvard.edu/ciao/threads/index.html} at the Chandra X-ray Center (CXC).  All level-1 data were reprocessed to the level-2 stage using standard CIAO tasks. 

In order to remove the "gridded" appearance of the images and to avoid any aliasing effect, the standard Chandra data reduction pipeline randomizes the X-ray event positions detected within a given pixel. This randomization process could slightly degrade the source centering accuracy. We thus removed the pixel randomization by reprocessing the raw data according to instructions available at CXC and analyzed these data in addition to the usual  randomized sets. 

Both the {\em celldetect} and {\em wavdetect} CIAO packages were used for source detection and characterisation. The task {\em celldetect} uses a sliding cell of variable size to search for locations where the signal to noise ratio of a candidate source is larger than a given threshold. The position of the source is then derived from the center of gravity of the X-ray events found in the detect cell. In {\em wavdetect}, source detection is performed by convolving the X-ray image with wavelets of various spatial extents. The position of the source is eventually derived using the same algorithm (centroid) as for {\em celldetect}\footnote{ http://cxc.harvard.edu/ciao/download/doc/detect\_manual/}. 

Exposure maps were created in order to avoid spurious detections along the edges of the CCDs and obtain more accurate positions for those sources located close to these edges. These maps were only used when running the {\em celldetect task}. Pixel scales of 1, 2 and 4 were used for {\em wavdetect}. These values are well suited to the detection of unresolved sources at moderate off-axis distances. We only considered sources detected in the central CCDs, chips I0 to I3 (16.9\arcmin\ by 16.9\arcmin) and S2 and S3 (8.3\arcmin\ by 16.9\arcmin) for ACIS-I and ACIS-S respectively.

The final relative astrometric accuracy of the reference frames depends on the number of sources common to the two observations and on the quality of the determination of their positions. We thus tried five different source detection thresholds, defined in {\em celldetect} as the significance of the source expressed in Gaussian sigma and in {\em wavdetect} as the probability of a false detection at any given pixel.  These thresholds ranged from 1.5 to 2.75 for {\em celldetect} and from 10$^{-8}$ up to a level of 5$\times$10$^{-5}$ for {\em wavdetect}. Testing several different thresholds allows to find the best compromise between the number of common sources and the mean positional errors, both rising with increasing detection sensitivity. The largest threshold value in {\em wavdetect} implies that about 40 false sources per CCD chip enter the source list. However, since we only consider sources detected at both epochs and located at a maximum distance of one to three arcsec, the actual probability that the source common to both observations is spurious is very low. Best positions are obtained in the energy range for which the signal to noise ratio of the reference sources is strongest. This range depends on the relative energy distribution of these sources and of the diffuse extragalactic and instrumental backgrounds. For the neutron star, measured in the same conditions as the reference sources, the choice of the energy range is much less critical since the object is brighter than any of the background sources. We decided to test energy bands 0.3-10\,keV, 0.3-5\,keV, 0.3-3\,keV, 0.5-5\,keV and 0.5-2\,keV.

\subsection{Matching reference frames}

The absolute astrometric frame derived independently for each Chandra observation is already very accurate since the 90\% confidence radius is estimated to be of 0.6\arcsec\ close to aimpoint (Chandra X-ray Center). However, reaching the ultimate precision on differential astrometry requires to correct for the remaining relative random attitude errors affecting the observations at the two epochs. We thus carried out a relative boresight correction by allowing the reconstructed equatorial positions of the second epoch to be slightly shifted in right ascension and declination and subjected to a small rotation around the aimpoint with respect to first epoch. We searched for translations of $\pm$ 1\arcsec\ in each direction with steps of 10\,mas and for a rotation of up to $\pm$ 0.1\,deg with steps of 5$\times$10$^{-3}$\,deg. The best offset and rotation angles were estimated using a maximum likelihood scheme, which for a bivariate normal distribution is equivalent to minimize the quantity

\begin{equation}
Q\ = \sum_{i = 1}^{n} \left( \frac{\Delta_{\alpha\,i}^{2}}{\sigma_{\Delta\alpha\,i}^{2}} +  \frac{\Delta_{\delta\,i}^{2}}{\sigma_{\Delta\delta\,i}^{2}} \right) 
\end{equation}
where $\Delta_{\alpha\,i}$ and $\Delta_{\delta\,i}$ are the distances in right ascension and declination between the positions of source $i$ at epoch one and two after transformation. The errors on the corresponding distances are $\sigma_{\Delta\alpha\,i}$ and  $\sigma_{\Delta\delta\,i}$ and are computed from the individual source uncertainties in right ascension and declination returned by the centering algorithms. The accuracy with which we can link the astrometric reference frames of the two epochs is given by

\begin{equation}
\sigma_{\rm frame} = \sqrt{ \frac{1}{n (n-1)} \sum_{i=1}^{n} (\Delta_{\alpha\,i}^{2} + \Delta_{\delta\,i}^{2})}
\end{equation}

\subsection{Systematic errors}

The ultimate accuracy with which Chandra can achieve relative astrometry is described in Chapter 5 of the Chandra Proposers' Observatory Guide. In this study, a sample of over 1100 sources with more than 50 counts detected in the framework of the Chandra Orion Ultra Deep Project \citep{getman05} was boresight aligned and cross-correlated with the 2MASS catalogue. After taking into account 2MASS positional errors, the distribution of the distances between ACIS-I and 2MASS entries indicated a systematic error with 90\% confidence value of 0.15\arcsec\ on X-ray positions corresponding to a $\sim$ 0.07\arcsec\ radial one sigma residual error. Therefore, although the {\em celldetect} and {\em wavdetect} algorithms can in principle center a relatively bright X-ray source on the CCD chip with a theoretical accuracy only restricted by signal to noise ratio, the ultimate positioning accuracy is limited by other systematic effects of unknown origin, but probably related to pixel to pixel sensitivity changes, intra pixel sensitivity variations, thermally induced misalignement, etc. We thus quadratically added a systematic error of $\sigma_{sys}$ = 70\,mas to all ACIS source positions, including those of the bright target isolated neutron star. 

For our pairs of observations, $\sigma_{\rm frame}$ is of the order of 100\,mas and thus dominates the error budget. The total error on the displacement of the central target is thus
\begin{equation}
\sigma_{INS} = \sqrt{ \sigma_{INS,1}^{2} + \sigma_{INS,2}^{2}
+ \sigma_{\rm frame}^{2} + 2\,\sigma_{sys}^{2}} 
\end{equation}

\section{Checking the actual accuracy achievable}

Only few attempts were done so far to measure proper motions in X-rays and the feasibility of such an endeavour is not often discussed in the general literature nor in the technical documentations. Note, however, that very recently \cite{kaplan2008} discussed to some extent the various errors entering astrometric ACIS-I observations and reached similar conclusions to ours. Before assessing the reality of the small displacement of an X-ray source on the sky one needs to envisage all potential sources of errors. This is best done by measuring in X-rays the motion of a source whose proper motion is accurately known from observations at other wavelength, e.g. radio or optical. Unfortunately, to our knowledge, at the time of writing this paper, the Chandra database does not contain repeated ACIS observations of high proper motion isolated neutron stars with properties such that they can be used as test data. The only possibilities left to test the error budget were then first, to perform extensive simulations of moving targets and second, to use repeated observations of non moving objects to test systematic errors. We examine these two options in the following subsections. 

\subsection{Simulations}

The goal of these simulations was twofold. First, to estimate the accuracy with which a small displacement of the INS could be detected using the ACIS-I and ACIS-S configurations, allowing for different aimpoint positions and systematic offsets of the reference frames. Second, to find the most efficient centering method and the best suited parameters, such as energy range, detection threshold and randomization state.

We used {\em MARX 4.2.1} to create fake event lists mimicking the 2002 and 2005 observations of the fields of \rxzerohuit\ and \rxzeroquatre . The simulated observations had the same duration as the original ones and in addition, the same starting times. This last constraint allows to account for the slow degradation of the quantum efficiency of the CCDs, especially at energies below 1\,keV. For each camera, the simulated sources were those common (within 3\arcsec) to the actual 2002 and 2005 observations as detected in the 0.5-5.0\,keV band and with a threshold of 10$^{-6}$ for \rxzerohuit\ and in the 0.5-2.0\,keV band with a threshold of 10$^{-5}$ for \rxzeroquatre. A total of 26 and 12 sources were created for ACIS-I and ACIS-S simulations respectively. Each simulated source had the same position on the sky and photon flux as the actual one. Since most of the background reference sources are too faint to constrain their energy distribution, we assumed that they were all of extragalactic origin and that their spectra were described by an absorbed power law of spectral index $\Gamma = 1.7$, undergoing a hydrogen column density of $N_{H} = 5.21 \times 10^{21}$ cm$^{-2}$ (for the field of \rxzerohuit) and $N_{H} = 1.07 \times 10^{20}$ cm$^{-2}$ (for the field of \rxzeroquatre). The total Galactic column densities were derived from the far infrared emission maps of Schlegel et al. (1998) available at the NASA/IPAC Extragalactic Database and applying the Predehl \& Schmitt (1995) relation between optical and X-ray absorptions. The energy distribution of the neutron stars were taken as blackbodies with $kT = 96$\,eV and $kT = 44$\,eV absorbed by \nh\ = 4$\times$10$^{19}$ cm$^{-2}$ and \nh\ =1$\times$10$^{20}$ cm$^{-2}$ for \rxzerohuit\ and \rxzeroquatre\ respectively. Input count spectra were generated using {\em XSPEC 12.2.1}. We created a diffuse background using the observed deep blank fields provided by the Chandra calibration database, by re-projecting them to the tangent plane of the observations and for each simulation by randomly picking background events so as to reproduce the level of background noise of the actual observation in the 0.5-5.0\,keV energy band. Finally, sources and background were merged into a single event file.

\begin{table*}[ht]
\caption{Simulation parameters. $\Delta$ RA and $\Delta$ Dec are the displacements applied to the aimpoint or to position of the neutron star in the second observation with respect to those in the first one.}
\label{simpar}       
\begin{tabular}{lclccccc}
\hline\noalign{\smallskip}

Run  &	Epoch  & Aimpoint    &$\Delta$ RA    &$\Delta$ Dec& Displacement &$\Delta$ RA&$\Delta$ Dec\\
     &         & offset   & aimpoint & aimpoint& target & target & target \\    
     &	       &  (pixel) &(arcsec) &(arcsec) & (arcsec) &(arcsec) &(arcsec)\\ 
\hline      
R1   & 	2002   &   Nominal   & 0.000 & 0.000 &&&\\
     &	2005   &   Nominal   & 0.000 & 0.000 & 1.00   &-0.707 &-0.707\\
R2$^{*}$   &	2002   &   Nominal   & 0.000 & 0.000 &&&\\
     &  2005   &   Nominal   & 0.000 & 0.000 & 1.00   &-0.707 &-0.707\\
R3   &  2002   &   0.25      &-0.123 & 0.000 &&&\\
     &  2005   &   0.25      &-0.123 & 0.000 & 1.00   &-0.707 &-0.707\\
R4   &  2002   &   0.50      & 0.000 &-0.246 &&&\\
     &  2005   &   0.50      & 0.000 &-0.246 & 1.00   &-0.707 &-0.707\\
R5   &  2002   &   0.67      & 0.232 & 0.232 &&&        \\
     &  2005   &   0.67      & 0.232 & 0.232 & 1.00   &-0.707 &-0.707\\
R6   &  2002   &   Nominal   & 0.000 & 0.000 &&&\\
     &  2005   &   Nominal   & 0.000 & 0.000 & 0.75   &-0.375 &-0.650\\
R7   &  2002   &   0.25      &-0.123 & 0.000 &&&\\
     &  2005   &   0.25      &-0.123 & 0.000 & 0.75   &-0.375 &-0.650\\
R8   &  2002   &   0.50      & 0.000 &-0.246 &&&\\
     &  2005   &   0.50      & 0.000 &-0.246 & 0.75   &-0.375 &-0.650\\
R9   &  2002   &   0.67      & 0.232 & 0.232 &&&\\
     &  2005   &   0.67      & 0.232 & 0.232 & 0.75   &-0.375 &-0.650\\
R10  &  2002   &   Nominal   & 0.000 & 0.000 &&&\\
     &  2005   &   Nominal   & 0.000 & 0.000 & 0.34   &+0.232 &+0.253\\
R11  &  2002   &   0.25      &-0.123 & 0.000 &&&\\
     &  2005   &   0.25      &-0.123 & 0.000 & 0.34   &+0.232 &+0.253\\
R12  &  2002   &   0.50      & 0.000 &-0.246 &&&\\
     &  2005   &   0.50      & 0.000 &-0.246 & 0.34   &+0.232 &+0.253\\
R13  &  2002   &   0.67      & 0.232 & 0.232 &&&\\
     &  2005   &   0.67      & 0.232 & 0.232 & 0.34   &+0.232 &+0.253\\
     
\noalign{\smallskip}\hline

\end{tabular}

$^{*}$ For this run, all source positions in the second epoch were offset by a fraction of an arcsecond (see text)

\end{table*} 

We verified that simulated sources displayed a count histogram consistent with the expected Poisson distribution. The input flux of each simulated source was modified in order to account for the exposure map, in particular for those sources close to the edge of a CCD. We also checked that on average, the count rates of the simulated sources were identical to those of the observed ones.

Table \ref{simpar} lists the parameters of the 13 different settings used for the first and second epoch simulations. We applied three different offsets to the position of the central soft X-ray source representing the neutron star in the event file of the second epoch. A number of parameters were changed as well. The position of the aimpoint was moved by a fraction of a CCD pixel in order to produce slightly different distributions of the source PSF on the pixel grid. Note, however, that the slow Lissajous shaped dithering applied to Chandra pointing during observations should smear out PSF effects to a large extent. A systematic offset of 0.29 arcsec ($\pm$0.15 \arcsec and $\pm$0.25\arcsec in RA and DEC respectively) was applied to the 2005 coordinate system of run R2 to test our capability to correct for small shifts in the astrometric frames. Additionally, the randomization in pixel -- which is part of the standard processing in the Chandra pipeline -- was disabled in an extra sample of simulations in order to investigate the impact of this effect on the final positional accuracy. All these artificial observations were eventually processed by the same pipeline as the real data using the two CIAO detection algorithms, {\em celldetect} and {\em wavdetect}. 

None of the two algorithms, {\em celldetect} and {\em wavdetect} could recover all of the reference sources entered in the simulations, as a consequence of the background and photon counting noise introduced. At the maximum sensitivity used here, {\em celldetect} ($\sigma$ = 1.5) finds less sources than {\em wavdetect} (threshold of 5$\times$10$^{-5}$). Averaging over all simulations and energy bands used, {\em celldetect} finds on average 6.2 and 8.1 sources common to both epochs for the ACIS-S and ACIS-I simulations respectively while for {\em wavdetect} the figures are 7.0 and 14.8. 

We evaluated the relative merits of the two detection and positioning methods using as criterion the statistical error derived on relative frame matching between the two epochs. Table \ref{algos} lists the distribution of these frame errors for ACIS-I and ACIS-S, averaged over energy bands and detection thresholds for which the number of common sources was comprised between 7 to 10 for \rxzeroquatre\ and  12 to 16 for \rxzerohuit. None of the two algorithms proves a clear-cut advantage, their performances remain statistically consistent with each other. Removing pixel randomization while keeping all other parameters identical does slightly improve source centering accuracy. The effect remains however relatively small, at the level of a few percents. 

\begin{center}
\begin{table}[t]
\caption{Matching error (in arcsec) and detection method for ACIS-S and ACIS-I simulated observations.}
\centering
\label{algos}       
\begin{tabular}{lcccc}
\hline\noalign{\smallskip}

Field  & Camera & Method & $\sigma_{\rm frame}$ & rms \\\hline
\rxzeroquatre & ACIS-S& {\em celldetect} & 0.164  &     0.032\\
\rxzeroquatre & ACIS-S& {\em wavdetect} & 0.184   &    0.038\\
\rxzerohuit & ACIS-I& {\em celldetect} & 0.134    &   0.022\\
\rxzerohuit & ACIS-I& {\em wavdetect} & 0.128     &  0.019\\
\noalign{\smallskip}\hline
\end{tabular}
\end{table} 
\end{center}

Not unexpectedly, the energy band considered for source detection also affects to some extent the accuracy with which X-ray sources are positioned. Wide bands allow more sources to be detected, but introduce a higher background. For our simulations, the wider (0.3-10\,keV) and narrower (0.5-2\,keV) bands tend to yield slightly worse errors on positions. The small offsets ($\pm$ 0.15\arcsec\ and $\pm$ 0.25\arcsec) applied to the coordinate system of the second epoch of run R2 are also properly recovered. 

\begin{figure*}[ht]
\caption{Mean displacements in $\alpha$ and $ \delta$, averaged over the selected simulations (see text), compared to their input values (diamonds). Error bars show the rms of the measurements values. Black and red squares denote {\em celldetect} and {\em wavdetect} respectively.}
\begin{tabular}{cc}
\psfig{figure=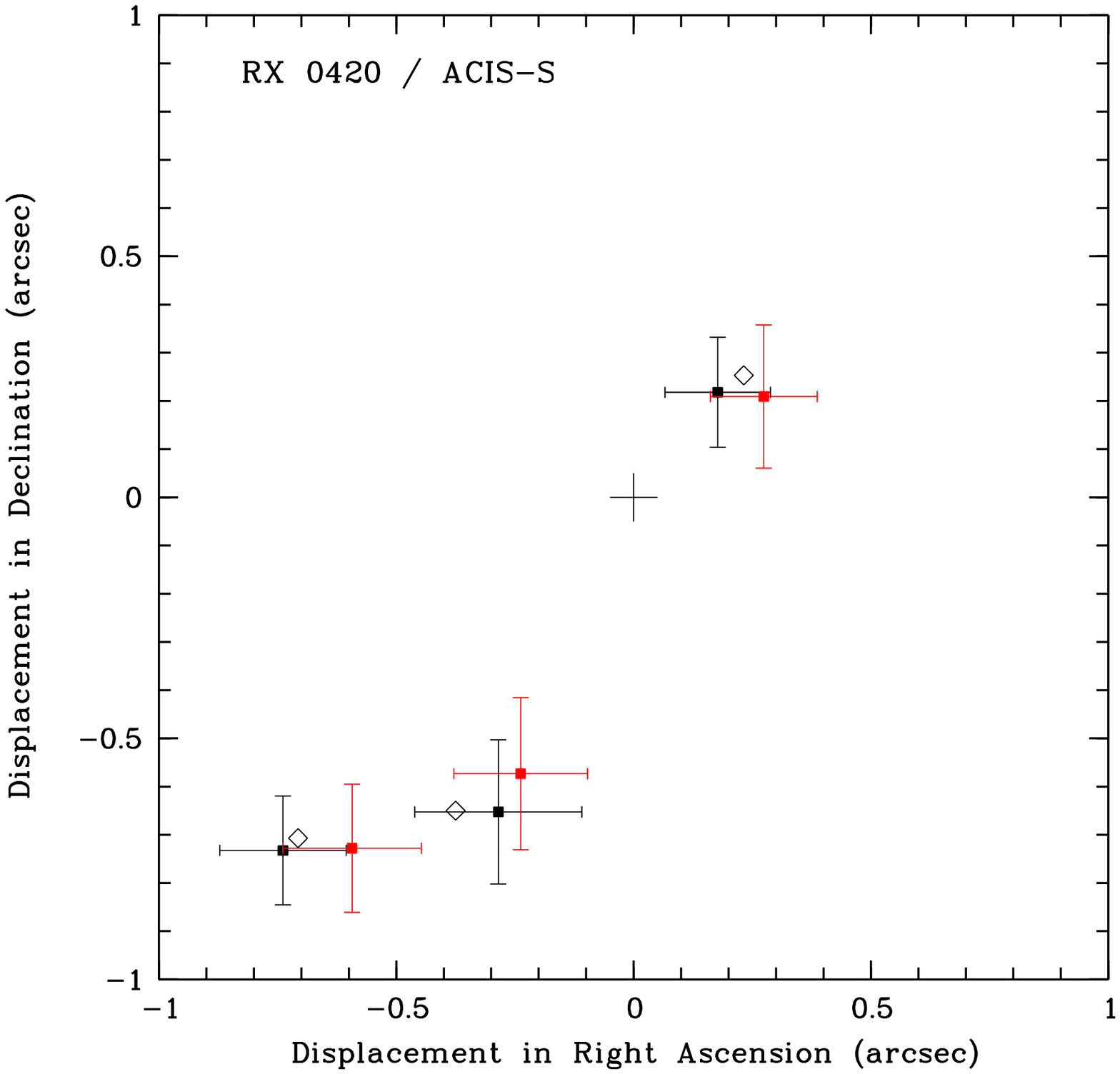,height=8.5cm,angle=-90,bbllx=45pt,bblly=74pt,bburx=570pt,bbury=600pt,clip=true} &
\psfig{figure=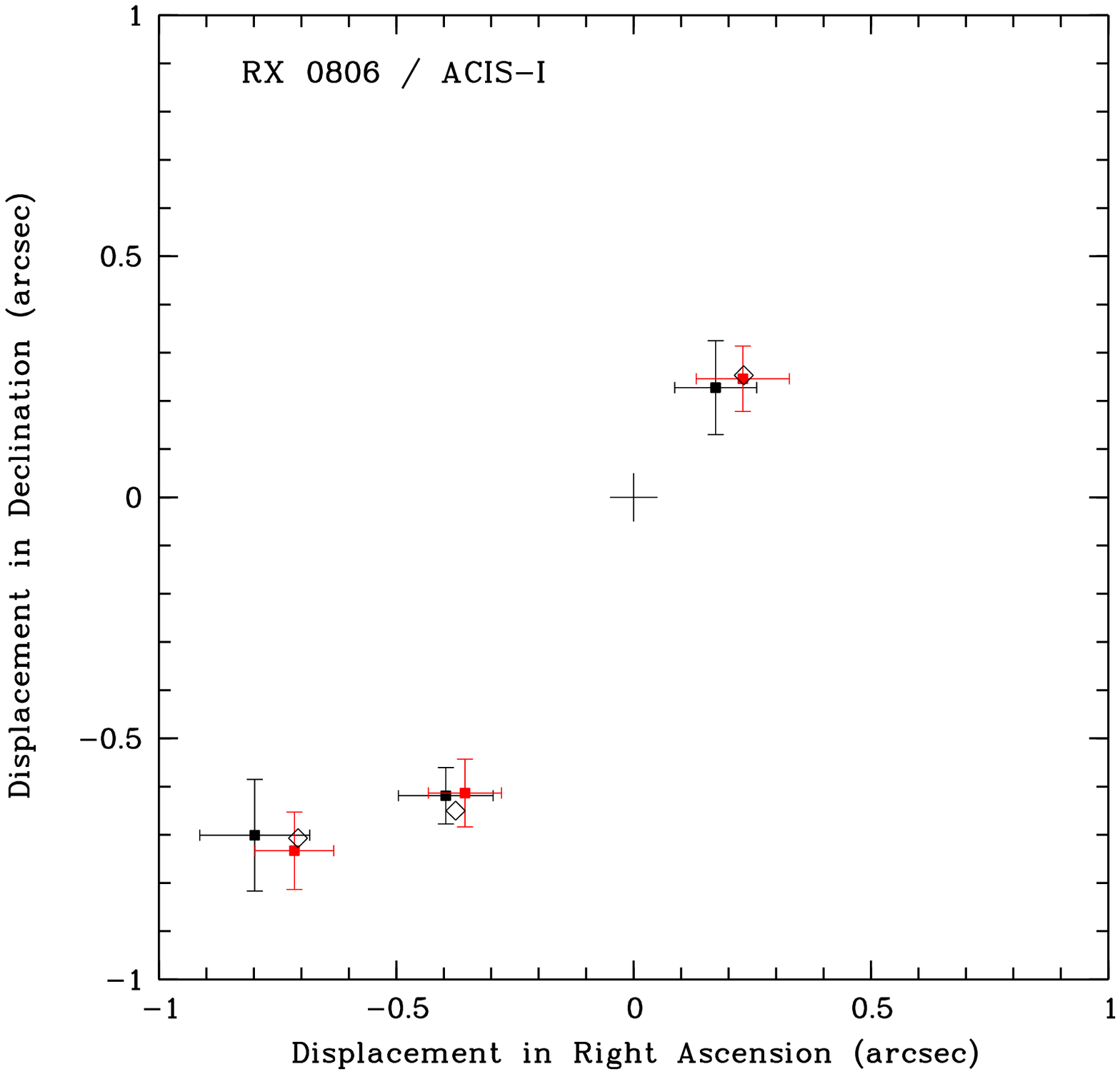,height=8.5cm,angle=-90,bbllx=45pt,bblly=74pt,bburx=570pt,bbury=600pt,clip=true} \\
\label{simdis}
\end{tabular}

\end{figure*}

We finally checked the accuracy with which the small motion of the simulated neutron star could be sized after frame matching. We show in Fig. \ref{simdis} the measured displacement of the neutron star compared to the one entered in the simulation. We applied here to the reductions a selection similar, although not exactly identical, to that applied to the real data. Namely, we impose the presence of at least 7 reference sources for ACIS-S and 12 for ACIS-I and we discard reductions yielding frame errors larger than the mean of all reductions. Randomized and de-randomized simulations are included. Error bars show the dispersion of the results from the selected reductions and simulations. Shifts as small as 0.34\arcsec\ are properly recovered with an accuracy of 0.138\arcsec\ and 0.075\arcsec\ for the ACIS-S and ACIS-I configurations respectively using the {\em wavdetect} algorithm.

\subsection{Repeated fields}

Our simulations gave us high confidence in the possibility to measure small displacements of unresolved X-ray sources with ACIS. The {\em MARX} ray tracing suite relies on the characteristics of the telescope and cameras as measured before launch in the laboratory and later evolved to account for actual flight performances. However, a number of unadvertised effects might introduce additional errors on the final relative astrometry recoverable in real observations. We thus searched the Chandra archive for repeated observations of extragalactic targets fulfilling similar conditions as ours in order to confirm that the relative positioning errors were indeed consistent with those estimated from simulations. Constraints were the existence of a rather bright unsaturated unresolved X-ray source with enough relatively faint reference sources around in pairs of ACIS images obtained with as far as possible similar roll angles and exposure times. Not surprisingly, very few observations match these conditions since only programmes aimed at long or short term monitoring are good candidates, the rest of the repetitions usually implying different modes or different instruments. We eventually found 3 pairs of ACIS-I and 3 pairs of ACIS-S observations suitable for the tests. We list in Table \ref{repeatedfields} the main properties of these fields.
These pairs of observations were processed with exactly the same pipeline as that used for our astrometric programme. In four instances, the source whose position was monitored was the original AGN target. For the two "Field" observations we had to select a relatively bright source (with a total number of counts larger than $\sim$ 40 in the shortest exposure) and located as close as possible to the aimpoint. As for our INSs observations, we estimated the displacement of the target source using the five combinations of detection thresholds and energy bands providing the lowest $\sigma_{\rm frame}$. We only discuss here the results of the {\em wavdetect} algorithm since {\em celldetect} yields quite similar frame errors and displacements. Removing pixel randomization does not necessarily improve frame matching. We list in Table \ref{repeatedresults} the measured displacements of the target source and their statistical significance. None of the extragalactic targets shows a significant motion on the sky since in all cases its significance remains below $\sim$1$\sigma$. The largest shift is seen on the LALA Cetus field target which is also the X-ray faintest of our test sources. In all other cases, the displacements of the bright source measured after matching the astrometric frames is smaller than 100\,mas. These results agree quite well with those of our simulations and give us further confidence in the astrometric capabilities of the Chandra Observatory. Interestingly, small shifts in the coordinates of the aimpoint and large changes in the roll angle do not seem to impact much the accuracy of the relative astrometry. 

\begin{center}
\begin{table*}[t]
\caption{Repeated Chandra observations used as test fields.}
\centering
\label{repeatedfields}       
\begin{tabular}{lcccccl}
\hline\noalign{\smallskip}
Target             & Observation& Camera & Mode   & Exposure & Roll angle & Target Used\\[3pt]
                   & date       &        &        & time (ks)& (deg)      &     \\[3pt]
\hline
LALA Cetus FIELD   & 2003-06-13 & ACIS-I & VFAINT & 160.1 & 121.5 & Source at \\
LALA Cetus FIELD   & 2003-06-15 & ACIS-I & VFAINT & 14.2  & 121.6 & 02:04:46.5s +05:04:04.4 \\

PKS0312-770        & 1999-09-08 & ACIS-I & VFAINT & 12.8 & 63.8 & PKS 0312-770\\
PKS0312-770        & 1999-09-08 & ACIS-I & VFAINT & 12.6 & 63.8 &\\

FIELD-142549+353248& 2002-04-16 & ACIS-I & VFAINT & 120.1 & 163.4  & Source at \\
FIELD-142549+353248& 2002-06-09 & ACIS-I & VFAINT & 58.5  & 219.9  &14:25:47.1  +35:39:54.8\\

3C 33              & 2005-11-08 & ACIS-S & FAINT  & 19.9  &  281.45 & 3C 33 \\
3C 33              & 2005-11-12 & ACIS-S & FAINT  & 19.9  &  281.45 & \\

3C 325             & 2005-04-14 & ACIS-S & VFAINT & 29.6  & 145.22 & 3C 325\\
3C 325             & 2005-04-17 & ACIS-S & VFAINT & 28.7  & 145.22 & \\

MG J0414+0534      & 2001-11-09 & ACIS-S & FAINT  & 28.4  & 57.35 & MG J0414+0534 \\
MG J0414+0534      & 2002-01-08 & ACIS-S & VFAINT & 96.7  & 295.24 & \\

\noalign{\smallskip}\hline
\end{tabular}
\end{table*} 
\end{center}
\begin{center}
\begin{table}[t]
\caption{Measured displacement $\Delta$ (mas) of the central source in repeated fields and its statistical significance.}
\centering
\label{repeatedresults}       
\begin{tabular}{lcccccl}
\hline\noalign{\smallskip}

Field              &  $\Delta$ & $\sigma$   &  $\Delta$   & $\sigma$  \\  
                   &\multicolumn{2}{c}{randomized} & \multicolumn{2}{c}{no randomization} \\ 
\hline		   
LALA Cetus Field   &  173        & 0.92        &   96        &  0.59 \\
PKS 0312-7707      &  100        & 0.88        &  112        &  1.02 \\
FIELD-142549+353248&   54        & 0.38        &   84        &  0.58 \\

3C 33              &   62        & 0.50        &   91        &  0.70 \\
3C 325             &   66        & 0.57        &   43        &  0.40 \\
MG J0414+0534      &  112        & 1.03        &   48        &  0.49 \\  
\noalign{\smallskip}\hline
\end{tabular}
\end{table} 
\end{center}

\section{Results}

Some reference sources showed rather large positional differences between the two epochs. Although these position offsets were not statistically significant considering the associated errors, we investigated how the inclusion of these sources in the transformation from one epoch to the other was impacting the value of the frame errors. Several test runs performed on the actual data showed that best frame errors were obtained when only sources located at a 
maximum distance of 1\arcsec\ were included in the transformation. Sources showing larger positional differences were in almost all cases located at large distances to the optical axis. In other instances, one of the two sources or both sources had a very small detection significance suggesting a spurious detection. We also tried both the {\em celldetect} and {\em wavdetect} source detection and centering tasks. For ACIS-I the error on the transformation delivered by {\em celldetect} was 10 to 70\% larger than that resulting from {\em wavdetect} positions. These results are at variance with the outcome of our simulated runs and of the repeated fields which showed that the two methods provided very comparable accuracies. In the case of ACIS-S (\rxzeroquatre), however, the two algorithms yielded similar results. We also note that the value of the displacement of the central source does not vary significantly with the task used. Therefore, in the following, we will only consider results obtained with the {\em wavdetect} task. 

\begin{figure*}[ht]
\caption{Accuracy of the astrometric matching (frame error in arcsec) as a function of the number of common sources plotted for different energy bands.}
\begin{tabular}{cc}
\psfig{figure=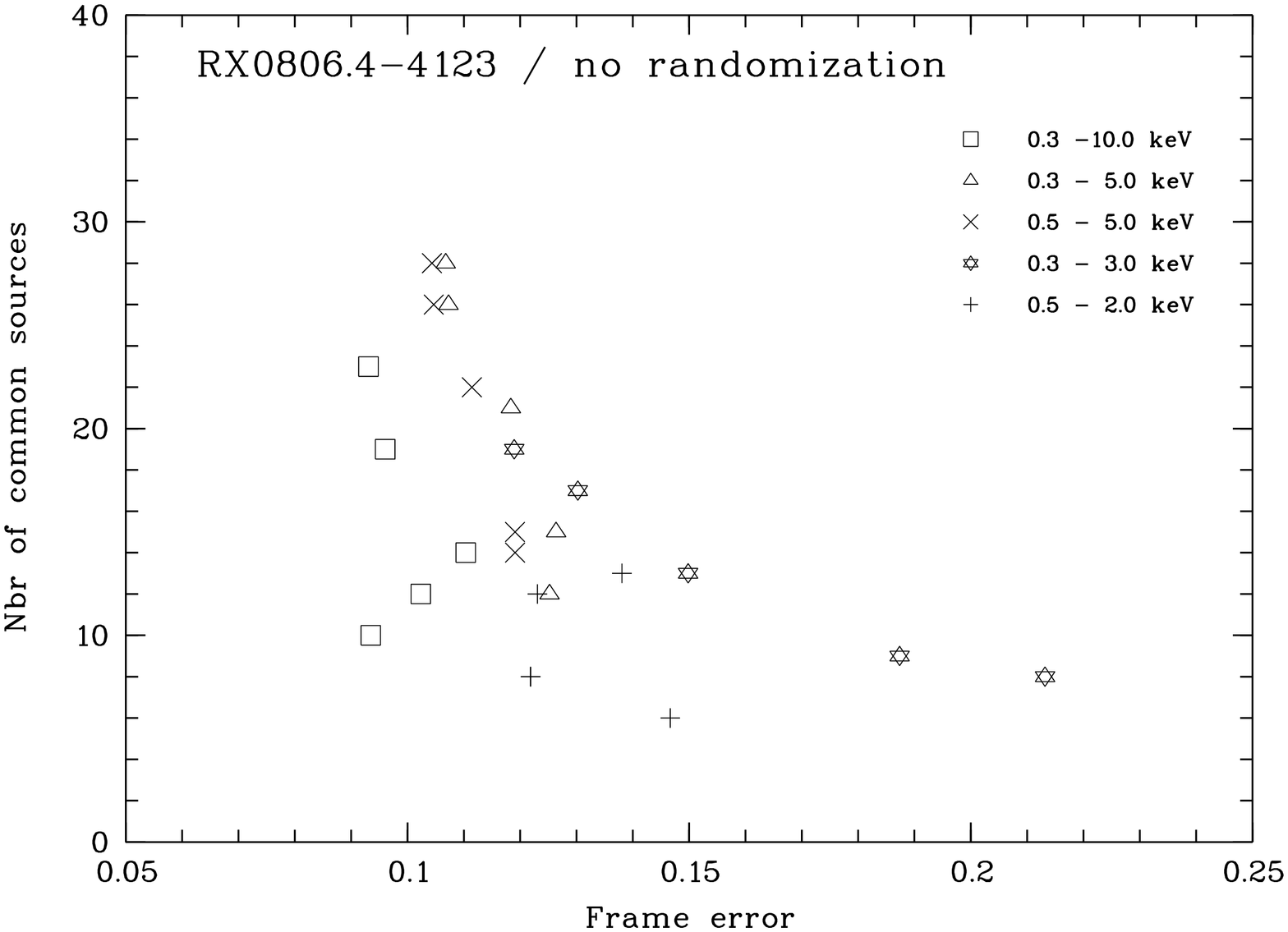,height=8.0cm,angle=-90,bbllx=5pt,bblly=74pt,bburx=600pt,bbury=790pt,clip=true} &
\psfig{figure=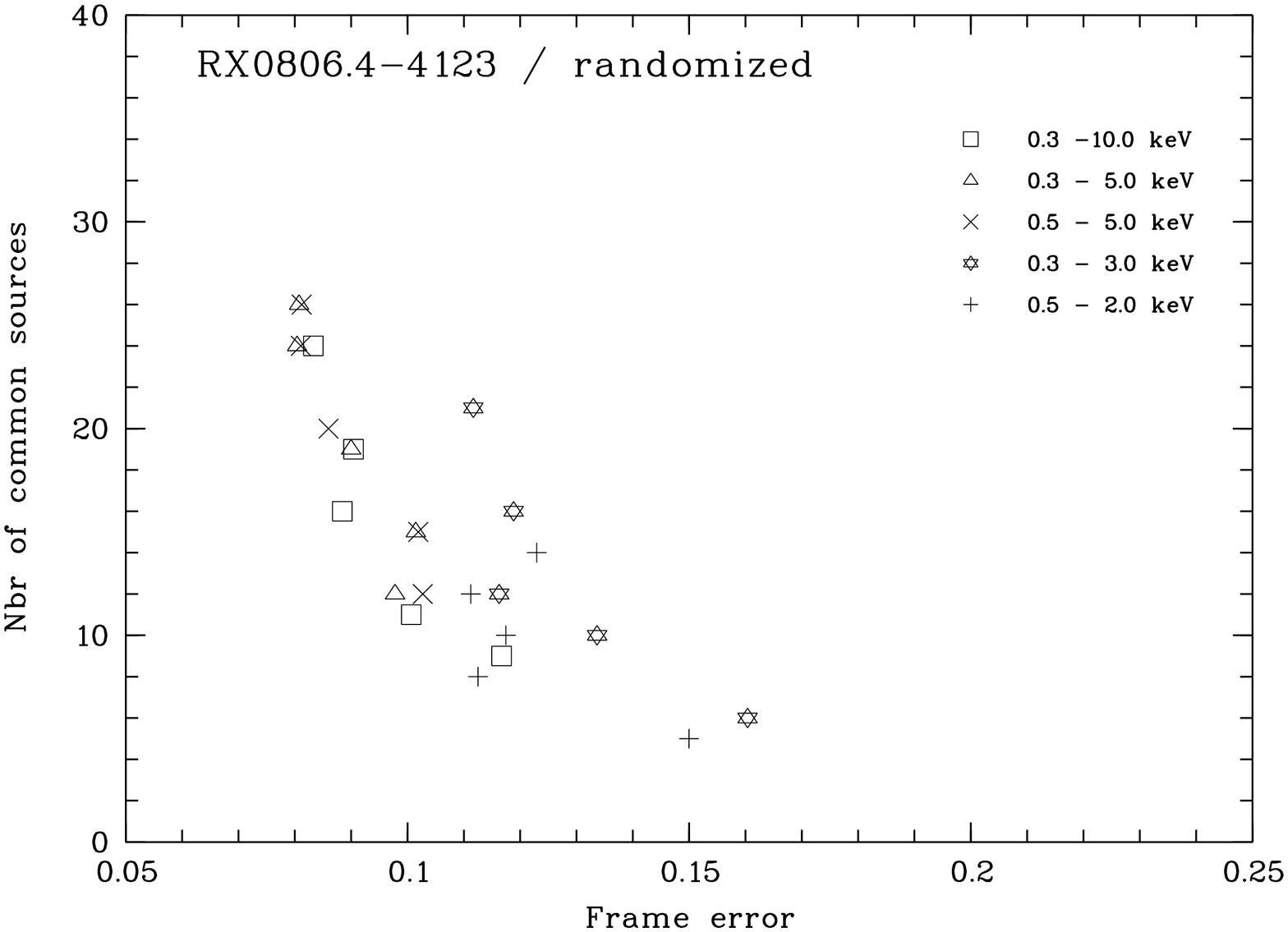,height=8.0cm,angle=-90,bbllx=5pt,bblly=74pt,bburx=600pt,bbury=790pt,clip=true} \\
\psfig{figure=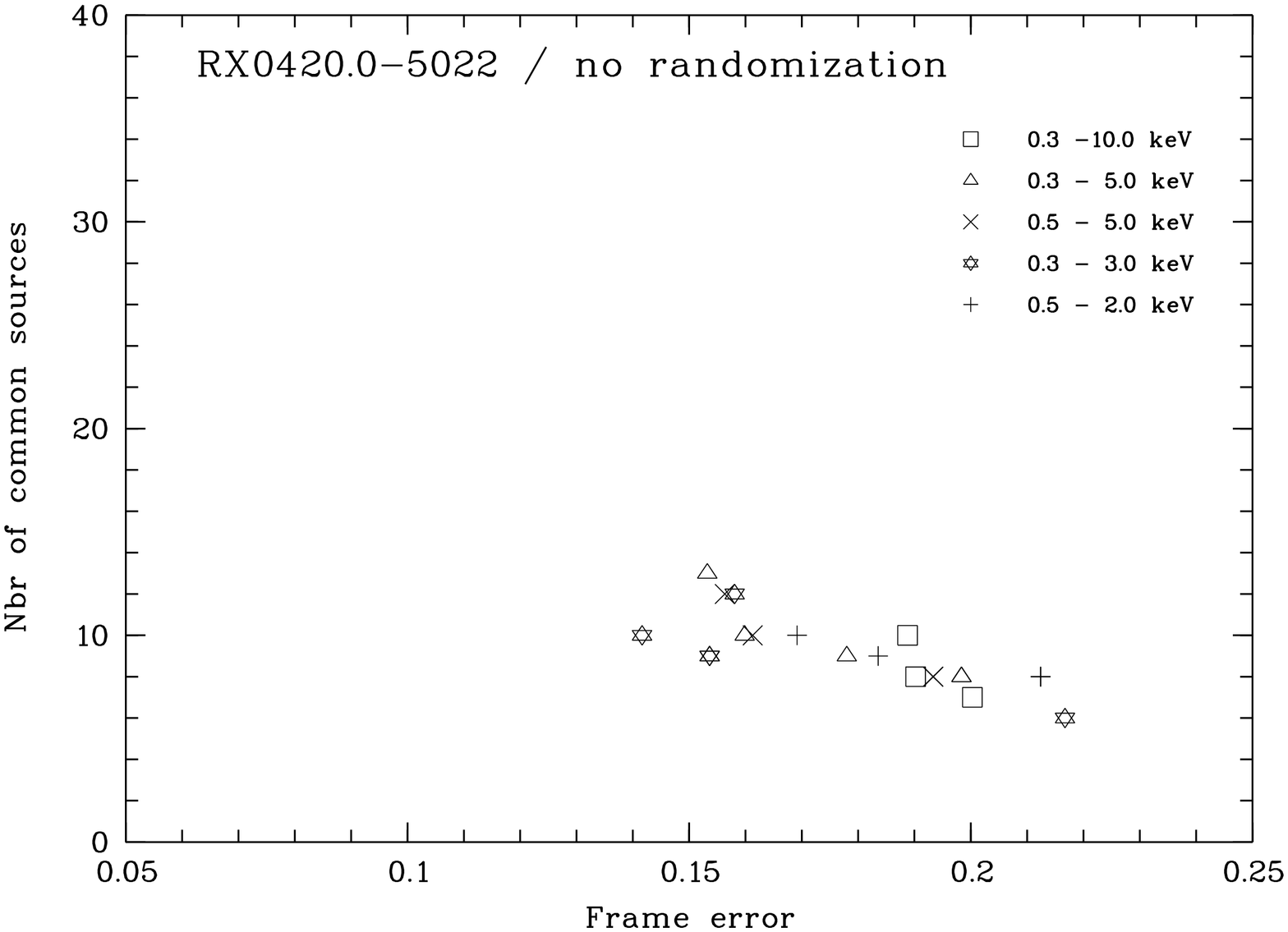,height=8.0cm,angle=-90,bbllx=5pt,bblly=74pt,bburx=600pt,bbury=790pt,clip=true} &
\psfig{figure=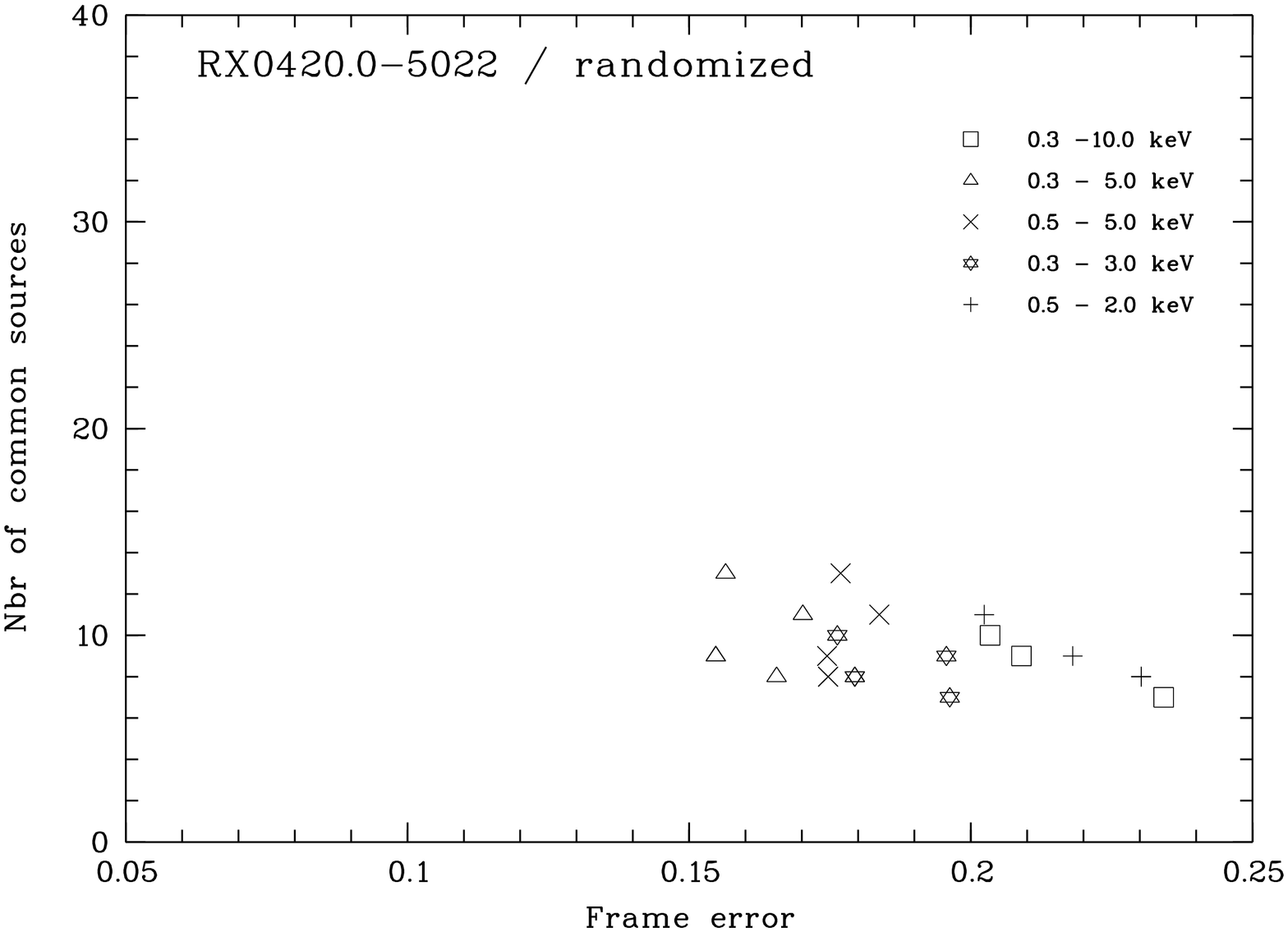,height=8.0cm,angle=-90,bbllx=5pt,bblly=74pt,bburx=600pt,bbury=790pt,clip=true} \\
\psfig{figure=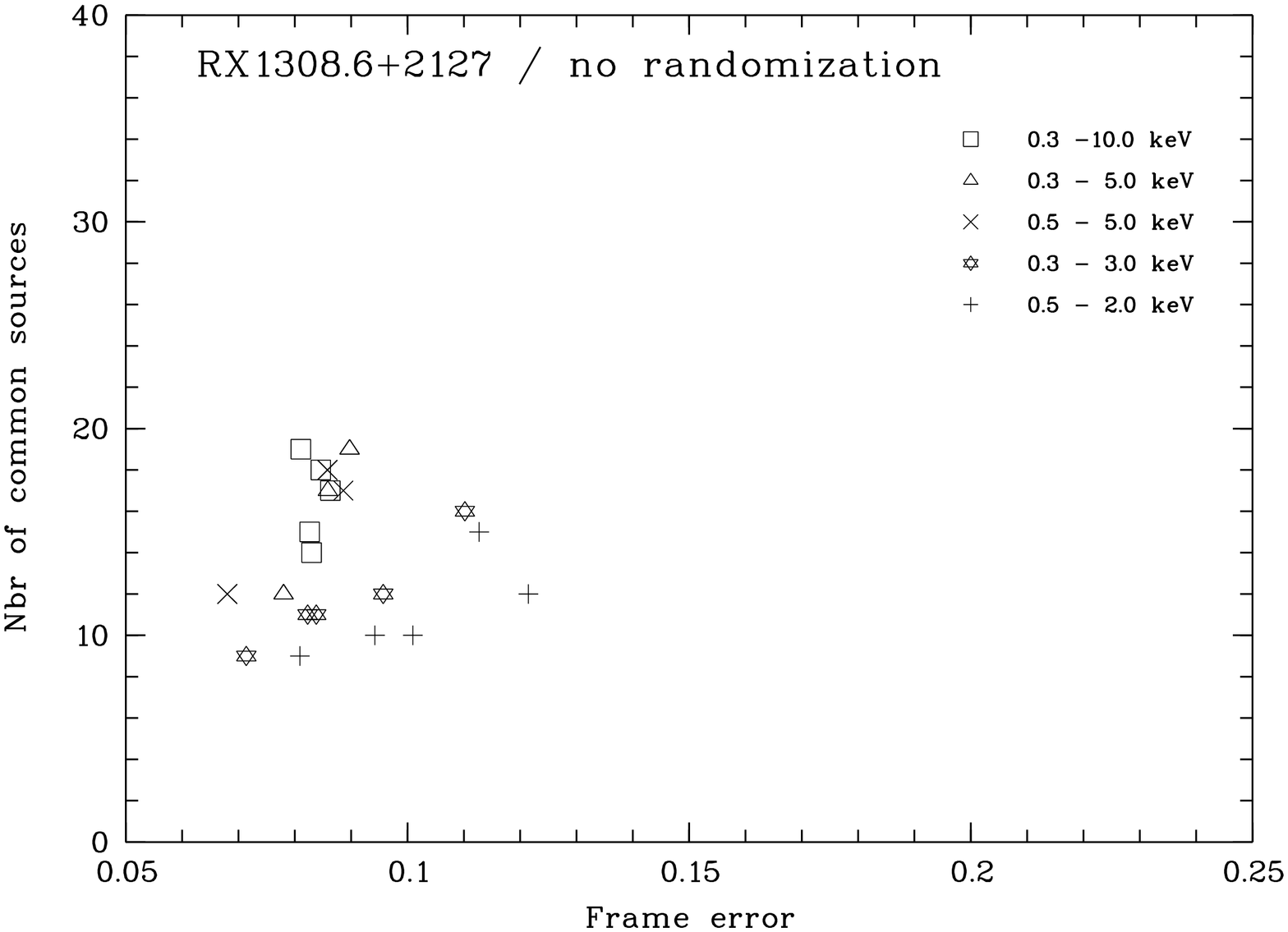,height=8.0cm,angle=-90,bbllx=5pt,bblly=74pt,bburx=600pt,bbury=790pt,clip=true} &
\psfig{figure=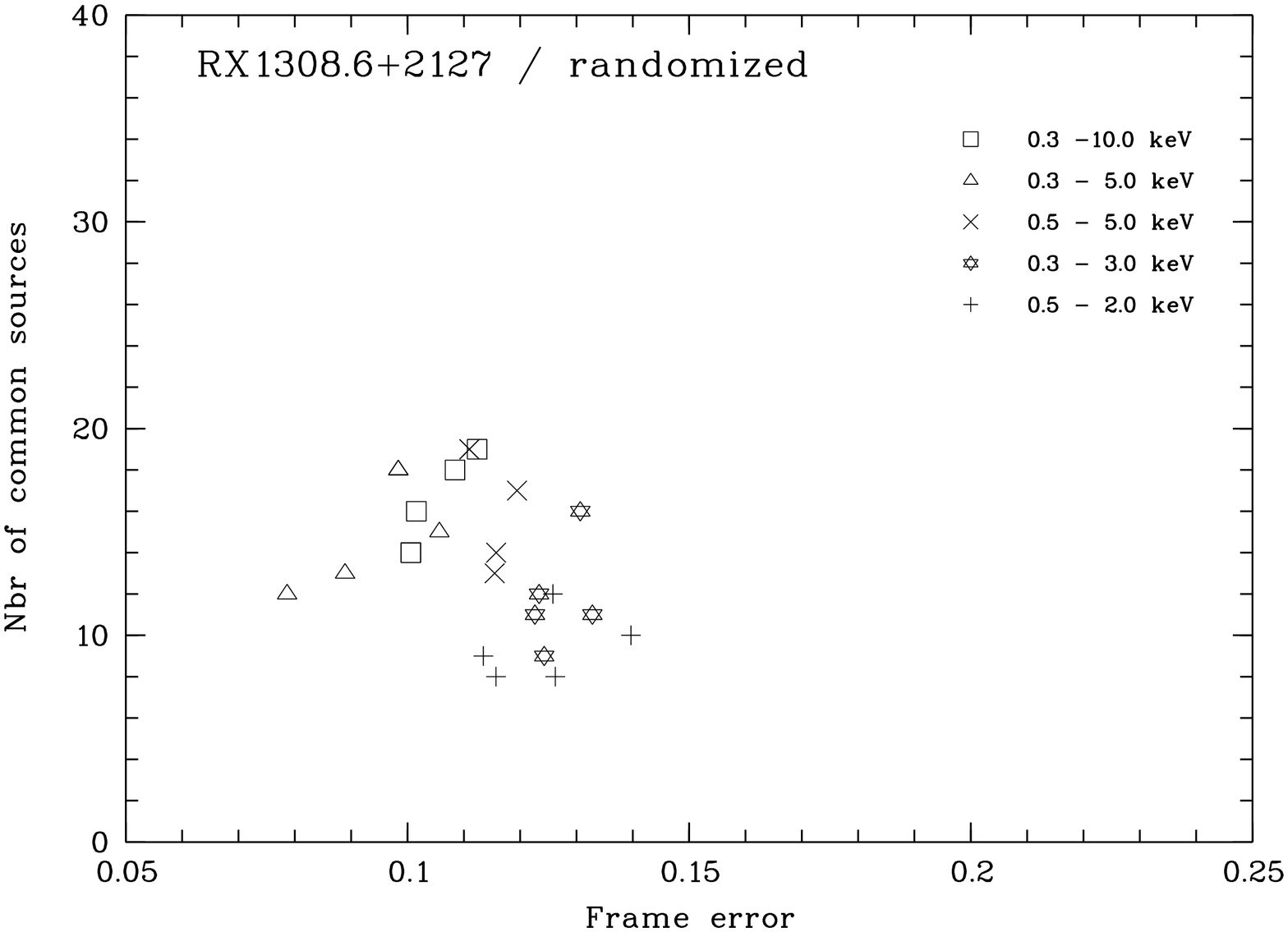,height=8.0cm,angle=-90,bbllx=5pt,bblly=74pt,bburx=600pt,bbury=790pt,clip=true} \\
\end{tabular}
\label{frameerrorncommon}
\end{figure*}

As for simulated data and repeated fields, the accuracy with which the two astrometric frames can be matched together depends on the randomization being applied or not, on the energy band used for source detection and on the value of the detection threshold. We show in Fig. \ref{frameerrorncommon} how the frame error varies with the number of sources common to both epochs. This number depends on source detection threshold and on energy band.  

Not unexpectedly, the randomization process tends to slightly deteriorate source positioning, although in the case of \rxzerohuit, the situation is inverted. Better frame errors are also obtained on the randomized data of some of the repeated fields studied above. In a similar manner, the energy band providing the best frame error slightly varies with target. In general, the restricted 0.3-5\,keV, 0.3-3.0\,keV or 0.5-5\,keV bands yield the smallest errors. This behaviour is consistent with the increasing ACIS background above 5\,keV and lower number of photons in the narrower 0.5-2\,keV band and is similar to that found in simulations. 

\begin{figure}
\caption{Mean number of detected sources (solid line) and number of sources common to the two observations within 1\arcsec (dashed line) in the 0.3-5\,keV band.}
\psfig{figure=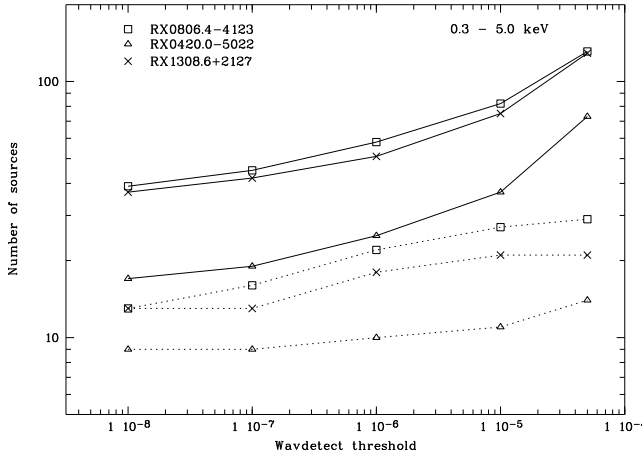,height=8.5cm,angle=-90,bbllx=5pt,bblly=74pt,bburx=600pt,bbury=790pt,clip=true} 
\label{Figncthres}
\end{figure}

Fig. \ref{Figncthres} shows the average number of sources detected in each observation and the number of sources common within 1\arcsec\ to both observations as a function of detection threshold. Data are presented here for the 0.3-5\,keV energy band. The threshold used in {\em wavdetect} represents the probability of a false detection at any given pixel of the image. Using wavelet scales of sizes 1, 2 and 4 pixels, over the central four ACIS-I CCD and central two ACIS-S chips, the expected number of false detections is of the order of $\sim$ 40 in ACIS-I at threshold = 10$^{-5}$. As expected, the number of detections dramatically rises with increasing thresholds reflecting the enhancement of the number of false sources. For \rxzerohuit\ and \rbsdouze\ (ACIS-I), the number of sources detected in each field increases by 24 when the threshold rises from 10$^{-6}$ to 10$^{-5}$. In contrast, the number of sources common to the two observations, which therefore have a high probability to be real increases much more slowly with threshold.

A significant number of high likelihood sources are not recovered from one observation to the next. At very low thresholds of 10$^{-7}$ or 10$^{-8}$ for which the number of spurious detections should be completely negligible, only one third of all sources are detected at both epochs. Selecting the brightest sources only, still half of them are not recovered from one observation to the other. This strongly suggests that a large fraction of extragalactic sources, mostly AGN, are significantly variable on a time scale of several years.

Overall, the density of field sources around the two INSs observed with ACIS-I are quite comparable and considering the very similar effective exposure times can be directly estimated from Fig. \ref{Figncthres}. The slightly higher number of detections around \rxzerohuit\ could be due to a contribution by additional stellar coronae which owing to the very low Galactic latitude of the target are expected to be well detectable. The number of ACIS-S detections is also very close to half that in ACIS-I in accord with the ratio of the field of view used for the two configurations. 

In order to be less sensitive to a particular configuration of reference sources and obtain more conservative errors, we averaged results obtained for the five (among a total of 25) combinations of detection thresholds and energy bands which yielded the lowest frame matching errors. Best matches of the astrometric reference frames were obtained with the parameters listed in Table \ref{matchpar}. 

\begin{center}
\begin{table}[t]
\caption{Measured frame offsets and rotation angle between two epochs. All distances are given in mas.}
\centering
\label{matchpar}       
\begin{tabular}{lccc}
\hline\noalign{\smallskip}
Target             & RA offset    & DEC offset  & Rotation \\[3pt]
                   & (mas)        & (mas)       & angle (deg)   \\[3pt] 
\hline
\rxzerohuit        & 156          &   -398      & -0.021    \\
\rxzeroquatre      & 248          &   132       & -0.034    \\
\rbsdouze          &  82          &   266       & -0.010    \\
\noalign{\smallskip}\hline
\end{tabular}
\end{table} 
\end{center}

We list in Table \ref{motions} the displacements and their associated errors measured between the two epochs and corrected for the values listed in Table \ref{matchpar}. The positions of \rxzerohuit \ and \rxzeroquatre\ in 2002 and 2005 are consistent within one Gaussian sigma. The two sigma upper limits on the proper motions of \rxzerohuit\ and \rxzeroquatre\ are 86\,\masyr\ and 123\,\masyr\, respectively. 

\begin{center}
\begin{table*}[t]
\caption{Measured displacements. All values given in mas.}
\centering
\label{motions}       
\begin{tabular}{lccccc}
\hline\noalign{\smallskip}
Target             & Target error & Frame Error & Total error   & Measured    & Significance \\[3pt]
                   &              &             &               & displacement& ($\sigma$)   \\[3pt] 
\hline
\rxzerohuit        & 12           &   81        &  128          & 117       &  0.91  \\
\rxzeroquatre      & 26           &   151       &  184          & 209       &  1.13  \\
\rbsdouze          & 8.5          &   73        &  123          & 1096      &  8.9   \\
\noalign{\smallskip}\hline
\end{tabular}
\end{table*} 
\end{center}

In contrast, a very significant displacement is detected for \rbsdouze, with a statistical significance of $\sim$ 9$\sigma$ and corresponds to a total proper motion of $\mu$ = 220\,$\pm$ 25\,\masyr. The magnitude and direction of the position move of \rbsdouze\ with respect to the background of extragalactic sources does not depend significantly on the randomization step being applied or not in the processing of the individual photons, nor does it depend on the number of reference sources considered. We also checked that the {\em celldetect} algorithm was yielding similar results. Expressed in J2000 equatorial coordinates, the proper motion vector is; 

\vskip 0.1cm
$\mu_{\alpha}$ $\cos$($\delta$) =  $-$207 $\pm$ 20 \masyr; 
$\mu_{\delta}$   =      84 $\pm$ 20 \masyr
\vskip 0.1cm

and corresponding proper motions in Galactic coordinates are: $\mu_{l}$ $\cos$($l$) = $-$122 \masyr\ and $\mu_{b}$ = 183 \masyr. The source is therefore moving toward larger Galactic latitudes as would be naively expected for a neutron star born in the Galactic plane. We note, however, that for such a high Galactic latitude object, the velocity perpendicular to the Galactic plane is dominated by the unknown radial component. \rbsdouze\ displays the second largest proper motion among the seven ROSAT discovered INSs, just after the X-ray brightest ROSAT INS \rxdixhuit. Several distance estimates are proposed in the literature. \cite{kaplan2002b} derive a distance of $\sim$\,670\,pc for \rbsdouze\ using the preliminary HST parallax of \rxdixhuit\ ($d$\,$\sim$\,140\,pc) and making the simplifying assumption that the optical emission arises from neutron star atmospheres with similar surface compositions and areas. The revised distance of 160\,pc for \rxdixhuit\ \citep{vankerkwijk2007} increases the estimated distance of \rbsdouze\ to $\sim$ 760\,pc. Under these assumptions, the distance predicted for \rxzerosept, $d\,\sim\,340\,$pc, is in fair agreement with that derived from the HST parallax, $d$ = 360$^{+170}_{-90}$\,pc \citep{kaplan2007}. Based on detailed X-ray light curve and SED modelling \cite{schwope2005} derived distances in the  range of 76 to 380\,pc. The short distance resulted for the extreme model which assumed small, isothermal hot spots on the cool stellar surface which remained undetected in X-rays. While this model explained the SED and the X-ray variability, it was regarded unlikely due to the unknown heating mechanism of the spots. The more realistic temperature model  based on  calculations by \cite{geppert2004} resulted in a comparative representation of the data but a much larger distance. This number is subject to uncertainties given the uncertain temperature profile and the unknown stellar parameters. The distance of 380\,pc was based on $R_{\rm ns} = 12$ km, $r_{\rm gr} = 4$, hence $R_\infty = 13.8$\,km. The growth of the integrated interstellar absorption with increasing distance has been used by \cite{posselt2007} to constrain the distance of several ROSAT discovered INSs. This method applied to INSs having astrometric parallaxes yields consistent values. However, the local absorption models which rests on the measurement of the Na I line equivalent width in hot or fast rotating stars has a too sparse coverage at high Galactic latitude to be used effectively and has to be replaced by a less accurate analytical description. In addition, the exact value of the photoelectric absorption derived from X-ray modelling somewhat depends on the number and position of the low energy shallow absorption lines. Fitting the XMM-Newton spectra of \rbsdouze\ with two low energy lines gives \nh\ $\sim$ 1.2-1.8$\times$10$^{20}$cm$^{-2}$ \citep{schwope2007}. This \nh\ value suggests a distance of at least $\sim$ 500\,pc \citep{posselt2007}. In this respect, the very short distance of 76\,pc proposed by \cite{schwope2005} seems incompatible with the observed photoelectric absorption. The best guess distance estimate of \rbsdouze\ is thus probably in the range of $\sim$400 to $\sim$800\,pc. We note that the proper motions published in \cite{motch2007} and \cite{motch2008} derived from a less extensive analysis are fully consistent with those reported here. 

\section{The possible age and birth place of \rbsdouze}

The accuracy with which Chandra is able to measure proper motion vectors does not exactly compare with what can be achieved using ground based telescopes or the HST. Nevertheless, the present determination turns out to be quite useful to estimate the possible origin of the neutron star and its likely dynamical age. 

We computed the past trajectories of the neutron star during the last 4 Myr assuming a $\pm$ 2$\sigma$ range in $\mu_{\alpha}$ and $\mu_{\delta}$, present distances from 50 to 1200\,pc and radial velocities in the interval of $\pm$ 700\,\kms. We corrected the observed proper motion vector for the displacement of the Sun with respect to the Local Standard of Rest. Considering the large velocity, young age and proximity of the neutron star and the relatively short time during which we follow it, we neglected effects due to Galactic gravitational potential and differential Galactic rotation. For most trajectories, these effects should be of $\la$ 1\% \citep{bienayme2006}. 

As already quoted above, a displacement toward higher Galactic latitude does not necessarily imply that \rbsdouze\ escapes from the Galactic plane since at $b$ = 83.082\degr, the unknown radial velocity dominates the component perpendicular to the Galactic plane. As a matter of fact, any radial velocity toward the Earth in excess of a few tens of \kms\ would imply that \rbsdouze\ is born at large distances from the Galactic plane and the vast majority of the trajectories escaping from the plane require receding radial velocities. 

We first investigated whether some backwards trajectories would cross any of the known nearby OB associations. For that purpose we compiled a merged list from the catalogues in \cite{humphreys1978}, \cite{ruprecht1981} and \cite{dezeeuw1999}, keeping for each association the most accurate information available, mainly the boundaries and mean distances determined by Hipparcos for the nearest groups. 

The number of past flight paths crossing any given OB association is low, less than $\sim$ 2\%. This reflects the small volume of the targeted associations compared to that sampled in total, which is necessarily large, owing to the substantial uncertainties on the proper motion vector, current distance and more importantly on the unknown radial velocity. In addition, only few nearby OB associations are counted in the $l$ $\sim$ 0\degr-20\degr\ range of Galactic longitude where \rbsdouze\ seems to originate from. The times at which backwards trajectories cross any given group of massive stars show a well defined peaked distribution with a small tail extending to the oldest ages considered in our study (4\,Myr) caused by few very low radial velocity trajectories. Hereinafter, current distances ($d_0$) and radial velocities ($V_r$) will refer to their mean over all trajectories comprised within one full width at half maximum of the peak of the age distribution. Our analysis shows that only three OB associations (see Table \ref{matchob}) could have given birth to \rbsdouze\ if we assume that the current distance to the neutron star is less than $\sim$ 900\,pc. Because of its proximity ($d$ = 145\,pc), the Upper Scorpius OB association ($l$ = -17\degr,0\degr; $b$ = 10\degr,30\degr) exhibits a significant motion which has to be taken into account when computing intersection times. The Upper Scorpius moving group is part of the general Sco OB2 association. Its past position was computed using group proper motion and radial velocity derived from Hipparcos and listed in \cite{dezeeuw1999}. According to \cite{reichen1990}, the Scutum OB2 association is likely to be the superposition on the line of sight of two disctinct groups, one located at 510\,pc and another at 1170\,pc.
Intersections with more remote ($d$ $>$ 1.5\,kpc) OB associations such as Sgr OB1, Sct OB3, Sgr OB7, Ser OB2, Sgr OB6, Ser OB1, Sgr OB4, Sgr OB5 all imply unrealistic present distances to the source larger than $d_0$ = 900\,pc with average crossing times of the order of 1.5  to 2.2 Myr. 
\begin{center}
\begin{table}[t]
\caption{Possible birth places of \rbsdouze\ listed with the present distance 
of the respective OB association and corresponding neutron star flight 
times, current distance and radial velocity ranges.}
\centering
\label{matchob}       
\begin{tabular}{lcccc}
\hline\noalign{\smallskip}
OB assoc & Distance & Age range    &   $d_0$  &  $V_r$ \\    
         &  (pc)    &   (Myr)      &  (pc) & (km/s) \\
\hline	  
UpperSco & 145	    &0.55$\pm$0.25 &  260$\pm$50 & -430$\pm$180\\  
SctOB2 A & 510	    &0.90$\pm$0.15 &  535$\pm$53 & -550$\pm$95 \\  
SctOB2 B & 1170	    &1.38$\pm$0.26 &  800$\pm$90 & -520$\pm$110\\  
\noalign{\smallskip}\hline
\end{tabular}
\end{table} 
\end{center}

However, a significant number of early type stars stars are found far from identified OB associations. The relative fraction of these field stars is not very well constrained and depends somehow on the accuracy of the photometric distances used and on the definition of the chosen boundaries for the associations. According to \cite{garmany1994}, the frequency depends on spectral type, the fraction of early type field stars varying from $\sim$ 40\% for stars hotter than B2 to 25\% for O stars only. The majority of field objects are runaway stars identified from their high velocities \citep{blaauw1961,stone1991} while the rest cannot be related to any group and might therefore be true field stars born outside clusters. Some of the runaway stars can convincingly be associated to the birth of radio pulsars \citep{hoogerwerf2001} and were thus kicked off their parent associations in the supernova explosion in a massive binary. Alternatively, the existence of many binary runaway stars \citep{gies1986} brings support to the dynamical ejection scenario in which early gravitational interaction with one or more cluster stars is responsible for the jump in velocity \citep{poveda1967}. 

In a second step, we thus relaxed the condition of a birth place close to a catalogued OB association and only imposed an origin in the supernova explosion of a field star assumed to be located within the Galactic plane.

\begin{figure}
\caption{Histograms of the crossing times of the Galactic plane for various assumed present distances to the source.}
\psfig{figure=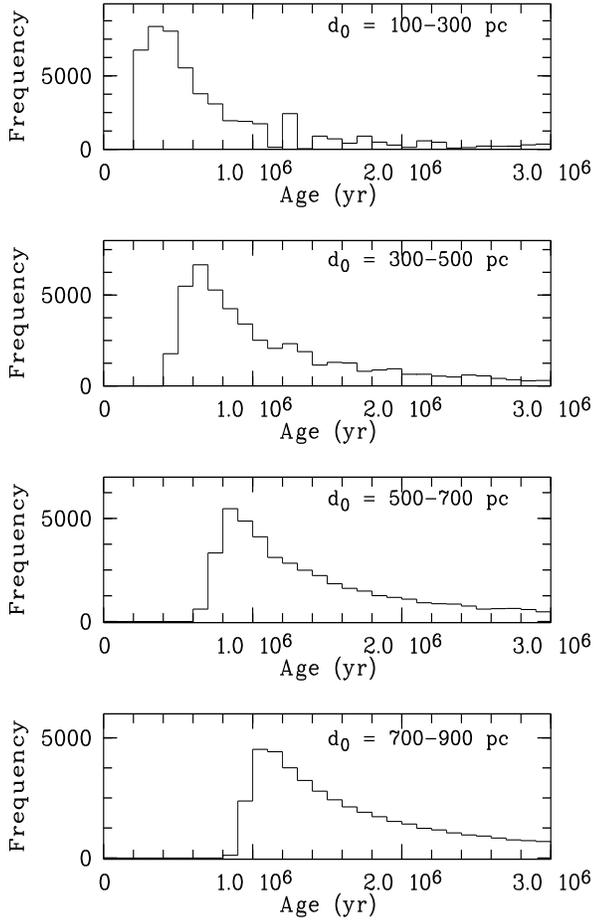,width=8.5cm,height=13cm,angle=0,bbllx=5pt,bblly=100pt,bburx=520pt,bbury=800pt,clip=true} 
\label{plot_his_gpp_approaches}
\end{figure}

Fig. \ref{plot_his_gpp_approaches} shows the histograms of the ages at which the backwards trajectories intercept the Galactic plane for different ranges of assumed present distances to the source. In a similar manner as for trajectories impacting OB associations, for a given range of current distances, the age histogram of all trajectories crossing the Galactic plane displays a clear peak with a reduced tail extending to longer times. Although ages older than 2 Myr appear possible for some parameters, the most probable ages peak at values increasing from 0.4$\pm$0.2 Myr for $d_0$ = 100-400\,pc to 1.3$\pm$0.3 Myr for $d_0$ = 800-1000\,pc (the interval of ages quoted here corresponds to the full width at half maximum of the peak of the histogram). For the most probable "intermediate" present distance range of 400 to 800\,pc, the age distribution peaks at 0.96 $\pm$ 0.30 Myr. Galactic longitudes at birth span a sector from approximately $l$ = -4\degr\ to $l$ = +24\degr. Old birth dates correspond to large present distances and more remote birth places ranging from 130\,pc ($d_0$ = 100-400\,pc) to 1600\,pc ($d_0$ = 800-1200\,pc). 

Note that the random position of the progenitor star with respect to the Galactic plane smears to a small extent the travel times computed above. The scale height of O to B5 stars belonging to the local Galactic disk is $\sim$ 40-60\,pc (see \cite{elias2006} and references therein). Taking this effect into account adds a scatter of $\sim$ 10$^{5}$\,yr to the ages shown in Fig. \ref{plot_his_gpp_approaches}.

The main parameter determining the age of \rbsdouze\ is its present distance which is therefore the most important source of uncertainty. A best guess $d_0$ of 400 to 800\,pc leaves as most likely birth place candidate the OB association Sct OB2 A and an age of 0.90$\pm$0.15 Myr. If the early type star from which \rbsdouze\ originates was a field object in the Galactic plane, then comparable ages of 0.96 $\pm$ 0.30 Myr appear as most probable. However, the much closer distances proposed by \cite{schwope2005} allow for a birth in the nearby Upper Scorpius part of the Sco OB2 complex 0.55$\pm$0.25 Myr ago. \rbsdouze\ could thus be like \rxdixhuit, also born in the Upper Scorpius \citep{walter2001}, and as \rxzerosept\ likely born in the Trumpler 10 or Vela OB2 associations \citep{motch2003,kaplan2007}, another product of the Gould Belt. 

\section{Discussion}

The proper motion of \rbsdouze\ is the second largest of all ROSAT discovered INSs (see Table \ref{propermotions}). However, owing to its plausible larger distance than other high proper motion members of the group, in particular that of the closest ROSAT INS \rxdixhuit, its tangential velocity V$_{\rm T}$ is probably considerable, in the range of 420 to 840\,\kms. Backwards trajectories crossing either a catalogued OB association or more generally the Galactic Plane also imply rather large radial receding velocities of the order of 550\,\kms , a value weakly dependent on the assumed present-day distance. The 3-D velocity of \rbsdouze\ might thus be of the order of 600 to 1000\,\kms. Such an enormous space velocity is nevertheless not unusual for neutron star standards. The radio pulsar \object{B1508+55} is a good example of a hyper-fast neutron star (V$_{\rm T}$ = 1083$^{+103}_{-90}$\,\kms) where both astrometric proper motion and trigonometric parallaxes were measured \citep{chatterjee2005}. The Guitar Nebula and its associated radio pulsar B2224+65 \citep{cordes1993} is another example of ultra high velocities (V$_{\rm T}$ = 1640\,($d$/1900\,pc)\,\kms). The high velocity of \rbsdouze\ precludes accretion of matter from the tenuous interstellar medium present at the large distance above the Galactic plane ($d_0$ $\ga$ 400\,pc) where the source is currently located (see e.g.  \cite{treves2000}). This INS is thus without doubt another young cooling neutron star.
  
Our Chandra observations of \rxzerohuit\ and \rxzeroquatre\ yield significant constraints on the tangential velocities of these two other sources. At a distance of 345\,pc derived from an analysis of the distribution of interstellar absorption on the line of sight \citep{posselt2007}, the transverse velocity of \rxzeroquatre\ already ranks among the lowest for the group. Moreover, the upper limit of 86\,\masyr\ obtained on the yearly displacement of \rxzerohuit\ is well below those displayed by the four sources for which a measurement exists. The probable small distance of the source ($d$ $\sim$ 235\,pc; \cite{posselt2007}) implies comparatively low transverse velocities of less than 96\,km\,s$^{-1}$. This brings the possibility that \rxzerohuit, being also the source located at lowest Galactic latitude among all ROSAT INSs, could power part of its X-ray emission through accretion of matter from the interstellar medium. This heating mechanism seems however unlikely since its X-ray spectrum, surface magnetic field and spin period do not differ much from those of other high velocity members of the group. Besides, the high magnetic field B $\sim$ 6.1 - 8.6 $\times$10$^{13}$\,G inferred by \cite{haberl2004} from the energy of the shallow absorption lines present in the soft X-ray spectrum and the rotation period of 11.37\,s would make accretion from a typical low to medium density interstellar medium impossible (see e.g. \cite{motch2000}). In addition, the lack of long term variability of the X-ray flux is at variance with what would be expected for an accreting neutron star.  

\begin{table}
\caption{Proper motions, distances and transverse velocities of the seven ROSAT discovered INSs. Upper limits are at the 2\,$\sigma$ level}
\begin{tabular}{lllll}
\noalign{\smallskip}
Name            &   $\ \ \ \ \mu$  &  Distance  	& V$_{\rm T}$   & Ref\\ 
                &   (\masyr) 	   &	(pc)		& (km\,s$^{-1}$)&  \\
\hline
\rxdixhuit      &   333$\pm$1	   & 161$^{+18}_{-14}$  & 254           & 1,2\\ 
\rxzerosept     &   107.8$\pm$1.2  & 361$^{+172}_{-88}$	& 184           & 3,4\\ 
\rxseize        & 155.0$\pm$3.1    &  $\sim$ 390	& 286           & 5,6,7\\
\rbsdouze       & 220.3$\pm$25     & 400-800            & 417-835       & 8,7,9\\ 

\rbsdixsept     &  -               & $\sim$ 430         & -             & 7\\
\rxzerohuit     & $\leq$ 86        & $\sim$ 235         & $<$ 96        & 7,9\\
\rxzeroquatre   & $\leq$123        & $\sim$ 345         & $<$ 200       & 7,9\\
\hline
\noalign{\smallskip}
\label{propermotions}
\end{tabular}

{References: 1) \cite{kaplan2002b}; 2) \cite{vankerkwijk2007}; 3) \cite{motch2003}; 4) \cite{kaplan2007}; 5) \cite{motch2005}; 6) \cite{zane2006}; 7) \cite{posselt2007}; 8) \cite{schwope2005}; 9) this work}

\end{table}

The present work adds three new measurements or upper limits and doubles the size of the sample of ROSAT INSs for which information on proper motion exists, leaving the seventh member, \rbsdixsept\ for further studies. It is now possible to better compare the distribution in transverse velocities of the ROSAT discovered INSs with that of radio pulsars. 

\begin{figure}

\psfig{figure=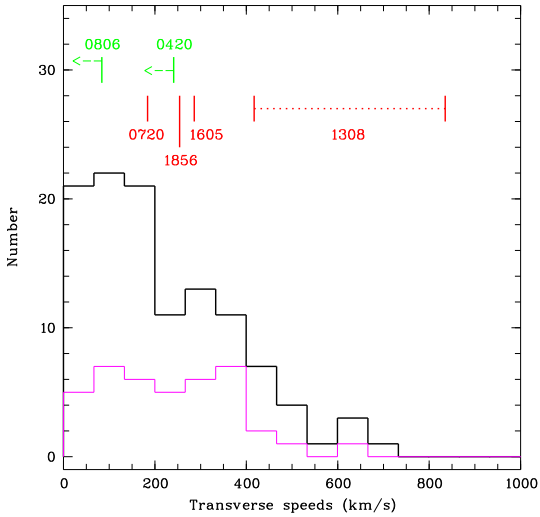,width=8cm,angle=-90,bbllx=5pt,bblly=74pt,bburx=600pt,bbury=620pt,clip=true}

\caption[]{Transverse speeds of ROSAT discovered INSs compared to those of non recycled radio pulsars. Data taken from Hobbs et al. (2005). The black thick line and the thin magenta line show the histograms of the entire population and of the young (age $\leq$ 3 $\times$10$^{6}$yr) respectively. Measured values and upper limits for the ROSAT INSs are shown on the top.}

\label{fig_histogramme_transverse_speeds-hobbspaper_lin}
\end{figure}

We plot on Fig. \ref{fig_histogramme_transverse_speeds-hobbspaper_lin} the distribution of the transverse velocities of non recycled radio pulsars, total and young (age $\leq$ 3$\times$10$^{6}$yr) populations as given in \cite{hobbs2005}. The positions of the six ROSAT INSs are also shown for comparison. ROSAT INSs exhibit 2-D velocity distributions consistent with those of radio pulsars. A simple Kolmogorov-Smirnov test gives a probability larger than 50\% that the distribution of the ROSAT INS tangential velocities is identical to that of the young (age $\leq$ 3$\times$10$^{6}$yr) radio pulsars. \rbsdouze\ fills in the large velocity tail of the radio pulsar distribution thus strengthening the similarity between the two populations. Within the limits enforced by small number statistics, it seems that the different birth conditions leading to the large magnetic fields and long spin periods of the ROSAT INSs are not related to a specific kick mechanism. This extends the apparent lack of dependency of radio pulsar properties on kick velocity (see e.g. \cite{wang2006}) toward larger B and spin period. 

\rbsdouze\ is the fifth brightest INS in the ROSAT PSPC band and might be the first to be born outside of the Gould Belt, either in one of the Sct OB associations or from a field OB star. This INS could thus mark the approximate flux boundary beyond which the production of other more distant OB associations or field stars becomes noticeable. We note that the 0.1-2.4\,keV PSPC count rate of 0.29\,cts\,s$^{-1}$ is roughly consistent with the population model of isolated cooling neutron star by \cite{posselt2008} which predicts that at PSPC count rates below $\sim$\,0.1\,cts\,s$^{-1}$ INSs located beyond the Gould Belt should dominate.  

Assuming that the distance to \rbsdouze\ is indeed smaller than 800\,pc yields ages of less than 0.80 Myr (for a birth in Upper Sco), 1.05 Myr (for a birth in Sct OB2 A) or 1.26 Myr for a birth from a field early type star. In contrast, the characteristic age derived from spin down $t_{sd}$ = 1.5\,Myr is somewhat longer. Similar and even greater disagreements between dynamical and spin-down ages are found for the two other ROSAT INSs for which $\dot{\rm P}$ was measured. In the case of \rxzerosept, the flight time from the most likely parent OB associations is in the range of 0.5 to 1 Myr for a characteristic age of 1.9 Myr \citep{kaplan2007}. Recently, \cite{vankerkwijk2008} reported the determination of the spin down of \rxdixhuit\ with again a huge discrepancy between the dynamical age of 0.4\,Myr, consistent with the estimated cooling age of $\sim$0.5\,Myr and the much longer spin down age of 4\,Myr. 

Several mechanisms have been put forward to explain this discrepancy (see \cite{kaplan2005b} and \cite{kaplan2005}). Among these are a long birth period of the order of several seconds and an early additional braking due to interaction with a fallback disc. The last mechanism envisaged is magnetic field decay. Originally proposed by \citep{heyl1998} it has been recently investigated in greater details by \cite{aguilera2008}. First, field decay provides through Joule heating an additional source of energy which accounts for a slower decrease of temperature with time. Second, the characteristic times derived from $\dot{P}$ then strongly overestimate the true age of the neutron star. Both effects converge to reconcile the relatively high polar temperatures observed in \rxzerosept\ and \rbsdouze\ with dynamical ages of up to 1\,Myr while explaining the discrepancy between dynamical and spin-down ages. In addition, the strongly mass dependent cooling efficiency makes the age temperature relation even more complex. If both mass and dynamical ages were known with enough accuracy, it could become possible, in principle, to estimate the strength of the magnetic field at birth and in the case of ROSAT discovered INSs find out whether these objects have evolved from a magnetar population (age $\sim$ 10$^ {6}$\,yr) or if middle-aged, were born with lower fields in the range of 10$^{13}$-10$^{14}$\,G \citep{aguilera2008}. The unusually large optical excess detected in \rbsdixsept\, \citep{zane2008} together with the large magnetic field evidenced from the soft X-ray absorption lines (B $\sim$ 10$^{14}$\,G; \citealt{zane2005}) hint at a close resemblance and perhaps evolutionary link between this ROSAT INS and magnetars. The magnetic field of \rbsdouze\ (B $\sim$ 4$\times$10$^{13}$\,G; \citealt{schwope2005}) ranks among the lowest of the ROSAT INSs \citep{haberl2007}. Decay from a typical magnetar field (B $>$ few 10$^{14}$\,G) would require an interval of time of the order of $\tau_{\rm Ohm} \,\sim$ 1\,Myr. A large present distance to the source could thus bring some support to a magnetar origin for \rbsdouze. However, the estimated relative birth rate of ROSAT INSs and magnetars seems to exclude that the entire population of ROSAT INSs evolves from former magnetars \citep{popov2006}

\section{Conclusions}

Chandra Observatory ACIS data obtained at a time interval of three and five years allow us to constrain the proper motion of \rxzerohuit\ and \rxzeroquatre\, with two sigma upper limits of $\mu$ $<$ 86\,\masyr\ and 123\,\masyr\, respectively. We measure the very significant displacement of \rbsdouze\ implying $\mu$ = 220.3 $\pm$ 25\,\masyr. The present work doubles the number of radio-quiet thermally emitting neutron stars for which information on proper motion exists. Extensive simulations carried out with {\em MARX} and analysis of fields observed several times by Chandra confirm the feasibility of measuring proper motions at X-ray wavelengths with an accuracy approaching that of optical ground based observations. 

The likely transverse velocity of \rbsdouze\ is large, of the order of 400 to 800\,\kms\ and could well be the largest of the six ROSAT discovered INSs for which a measurement of the proper motion and an estimate of the distance both exist. The receding radial velocities required for a birth close to the Galactic plane are of the order of 500\,\kms\ suggesting space velocities as high as 1,000\,\kms. \rbsdouze\ is thus another example of a young cooling radio-quiet nearby neutron star. Our upper limit on the displacement on the sky of \rxzerohuit\ hints at a slow relative velocity with respect to the interstellar medium. We argue, however, that accretion is unlikely to occur and does not significantly contribute to the X-ray emission.

Within the limits imposed by the small number statistics, the distribution in transverse velocity of the ROSAT INSs does not appear statistically different from that of the radio pulsars. This indicates that the mechanism leading to the strong magnetic fields and long periods typical of these objects is at first order independent from that controlling the kick velocity. 

The assumed present distance to \rbsdouze\ determines its possible travel time from nearby OB associations and from the Galactic plane in general. The analysis of all possible backwards trajectories suggest that for a present distance comprised between 400 and 800\,pc, the most likely birth OB association is the Scutum OB2 A group located at 510\,pc whereas a birth in the nearby Upper Scorpius part of the Sco OB2 association would be possible if the present distance to the source were much shorter ($d$ $\sim$ 260\,pc). We also find that travel times from the Galactic plane or from one of the candidate OB associations depend sensitively on the assumed current distance to the source and range from less than $\sim$ 1.2\,Myr for $d_0\,\sim\,400-800\,$pc to less than 0.6\,Myr for shorter distances. Although the still rather large uncertainties on the proper motion vector, current distance and the unknown radial velocity allow for longer flight times in some cases, the majority of the trajectories imply a kinematic age significantly smaller than the spin-down time of 1.5\,Myr. A similar discrepancy occurs for \rxzerosept\ and \rxdixhuit .

\begin{acknowledgements}

We acknowledge the use of Chandra X-ray Observatory data. The work of A.M.P. is supported by FAPESP (grant 04/04950-4), CAPES (grant BEX7812/05-7), Brazil, and the Observatory of Strasbourg (CNRS), France.

\end{acknowledgements}


\begin{thebibliography}{}

\bibitem[\protect\citeauthoryear{Aguilera et al.}{2008}]{aguilera2008} Aguilera, D.~N., Pons, 
J.~A., \& Miralles, J.~A.\ 2008, \apjl, 673, L167 


\bibitem[\protect\citeauthoryear{Bienaym{\'e} et al.}{2006}]{bienayme2006} Bienaym{\'e}, O., Soubiran, C., Mishenina, T.~V., Kovtyukh, V.~V., \& Siebert, A.\ 2006, \aap, 446, 933 

\bibitem[\protect\citeauthoryear{Blaauw}{1961}]{blaauw1961} Blaauw, A.\ 1961, \bain, 15, 
265 

\bibitem[\protect\citeauthoryear{Chatterjee et al.}{2005}]{chatterjee2005} Chatterjee, S., et 
al.\ 2005, \apjl, 630, L61 

\bibitem[\protect\citeauthoryear{Cordes et al.}{1993}]{cordes1993} Cordes, J.~M., Romani, 
R.~W., \& Lundgren, S.~C.\ 1993, \nat, 362, 133 

\bibitem[\protect\citeauthoryear{de Luca}{2008}]{deluca2008a} de Luca, A.\ 2008, 40 Years of Pulsars: Millisecond Pulsars, Magnetars and More, 983, 311 

\bibitem[\protect\citeauthoryear{de Luca et al.}{2008}]{deluca2008b} De Luca, A., Caraveo, 
P.~A., Esposito, P., \& Hurley, K.\ 2008, arXiv:0810.3804 

\bibitem[\protect\citeauthoryear{de Zeeuw et al.}{1999}]{dezeeuw1999} de Zeeuw, P.~T., Hoogerwerf, R., de Bruijne, J.~H.~J., Brown, A.~G.~A., \& Blaauw, A.\ 1999, \aj, 117, 354 

\bibitem[\protect\citeauthoryear{Elias et al.}{2006}]{elias2006} Elias, F., Cabrera-Ca{\~n}o, J., \& Alfaro, E.~J.\ 2006, \aj, 131, 2700 

\bibitem[\protect\citeauthoryear{Garmany}{1994}]{garmany1994} Garmany, C.~D.\ 1994, \pasp, 
106, 25 

\bibitem[\protect\citeauthoryear{Getman et al.}{2005}]{getman05}Getman, K.~V., et al.\ 
2005, \apjs, 160, 319 

\bibitem[\protect\citeauthoryear{Geppert et al.}{2004}]{geppert2004} Geppert, U., K{\"u}ker, M., \& Page, D.\ 2004, \aap, 426, 267 

\bibitem[\protect\citeauthoryear{Gies 
\& Bolton}{1986}]{gies1986} Gies, D.~R., \& Bolton, C.~T.\ 1986, \apjs, 61, 419 


\bibitem[\protect\citeauthoryear{Gotthelf \& Halpern}{2008}]{gotthelf2008} Gotthelf, E.~V., \& Halpern, J.~P.\ 2008, 40 Years of Pulsars: Millisecond Pulsars, Magnetars and More, 983, 320 

\bibitem[\protect\citeauthoryear{Haberl et al.}{1998}]{haberl1998} Haberl, F., Motch, C., 
\& Pietsch, W.\ 1998, Astronomische Nachrichten, 319, 97 

\bibitem[\protect\citeauthoryear{Haberl et al.}{1999}]{haberl1999} Haberl, F., Pietsch, W.,
\& Motch, C.\ 1999, \aap, 351, L53

\bibitem[\protect\citeauthoryear{Haberl \& Zavlin}{2002}]{haberl2002} Haberl, F., \& Zavlin, V.~E.\ 2002, \aap, 391, 571 

\bibitem[\protect\citeauthoryear{Haberl et al.}{2003}]{ha03}Haberl, F., Schwope, A.D., Hambaryan, V. et al. 2003, A\&A, 403, L19

\bibitem[\protect\citeauthoryear{Haberl et al.}{2004}]{haberl2004} Haberl, F., et al.\ 2004, \aap, 424, 635 

\bibitem[\protect\citeauthoryear{Haberl}{2007}]{haberl2007} Haberl, F.\ 2007, \apss, 308, 181
 
 
\bibitem[\protect\citeauthoryear{Heyl \& Kulkarni}{1998}]{heyl1998} Heyl, J.~S., \& Kulkarni, S.~R.\ 1998, \apjl, 506, L61 

\bibitem[\protect\citeauthoryear{Ho \& Lai}{2004}]{ho2004} Ho, W.~C.~G., \& Lai, D.\ 2004, \apj, 607, 420 

\bibitem[\protect\citeauthoryear{Hoogerwerf et al.}{2001}]{hoogerwerf2001} Hoogerwerf, R., de Bruijne, J.~H.~J., \& de Zeeuw, P.~T.\ 2001, \aap, 365, 49 

\bibitem[\protect\citeauthoryear{Hobbs et al.}{2005}]{hobbs2005} Hobbs, G., Lorimer, D.R., Lyne, A.G. et al. 2005, MNRAS, 360, 974

\bibitem[\protect\citeauthoryear{Hui et al.}{2006}]{hui2006} Hui, C.Y., Becker, W. 2006, A\&A, 457, L33

\bibitem[\protect\citeauthoryear{Humphreys}{1978}]{humphreys1978} Humphreys, R.~M.\ 1978, \apjs, 38, 309 

\bibitem[\protect\citeauthoryear{Kaplan et al.}{2002a}]{kaplan2002a} Kaplan, D.L., Kulkarni, S.R., van Kerkwijk, M.H. 2002a, ApJ, 579, L29

\bibitem[\protect\citeauthoryear{Kaplan et al.}{2002b}]{kaplan2002b} Kaplan, D.L., van Kerkwijk, M.H., Anderson, J. 2002b, ApJ, 571, 447

\bibitem[\protect\citeauthoryear{Kaplan \& van Kerkwijk}{2005a}]{kaplan2005b} Kaplan, D.L., van Kerkwijk, M.H. 2005a, ApJ, 628, L45

\bibitem[\protect\citeauthoryear{Kaplan \& van Kerkwijk}{2005b}]{kaplan2005} Kaplan, D.L., van Kerkwijk, M.H. 2005b, ApJ, 635, L65

\bibitem[\protect\citeauthoryear{Kaplan et al.}{2007}]{kaplan2007} Kaplan, D.~L., van Kerkwijk, M.~H., \& Anderson, J.\ 2007, \apj, 660, 1428 

\bibitem[\protect\citeauthoryear{Kaplan}{2008}]{kaplan2008} Kaplan, D.~L.\ 2008, 40 Years of Pulsars: Millisecond Pulsars, Magnetars and More, 983, 331 

\bibitem[\protect\citeauthoryear{Kaplan et al.}{2008}]{kaplan2008b} Kaplan, D.~L., Chatterjee, S., Gaensler, B.~M., Slane, P.~O.,  \& Hales, C.\ 2008, arXiv:0810.4184 

\bibitem[\protect\citeauthoryear{Kargaltsev \& Pavlov}{2008}]{kargaltsev2008} Kargaltsev, O., \& Pavlov, G.~G.\ 2008, 40 Years of Pulsars: Millisecond Pulsars, Magnetars and More, 983, 171 

\bibitem[\protect\citeauthoryear{Kondratiev et al.}{2008}]{kondra2008} Kondratiev, V.~I., 
Burgay, M., Possenti, A., McLaughlin, M.~A., Lorimer, D.~R., Turolla, R., Popov, S., 
\& Zane, S.\ 2008, 40 Years of Pulsars: Millisecond Pulsars, Magnetars and More, 983, 348 

\bibitem[\protect\citeauthoryear{Malofeev et al.}{2007}]{malo2007} Malofeev, V.~M., Malov, O.~I., \& Teplykh, D.~A.\ 2007, \apss, 308, 211 

\bibitem[\protect\citeauthoryear{McLaughlin et al.}{2006}]{mclaughlin2006} McLaughlin, M.~A., et al.\ 2006, \nat, 439, 817 

\bibitem[\protect\citeauthoryear{Motch et al.}{1997}]{motch1997} Motch, C., Guillout, P., Haberl, F., Pakull, M., Pietsch, W., \& Reinsch, K.\ 1997, \aap, 318, 111 

\bibitem[\protect\citeauthoryear{Motch}{2001}]{motch2000} Motch, C.\ 2001, X-ray 
Astronomy: Stellar Endpoints, AGN, and the Diffuse X-ray Background, 599, 
244 

\bibitem[\protect\citeauthoryear{Motch et al.}{2003}]{motch2003} Motch, C., Zavlin, V.~E., \& Haberl, F.\ 2003, \aap, 408, 323 

\bibitem[\protect\citeauthoryear{Motch et al.}{2005}]{motch2005} Motch, C., Sekiguchi, K., Haberl, F., Zavlin, V.~E., Schwope, A., \& Pakull, M.~W.\ 2005, \aap, 429, 257 

\bibitem[\protect\citeauthoryear{Motch et al.}{2007}]{motch2007} Motch, C., Pires, M.A., Haberl, F., Schwope, A.D. 2007, Ap\&SS, 308, 217

\bibitem[\protect\citeauthoryear{Motch et al.}{2008}]{motch2008} Motch, C., Pires, A.~M., 
Haberl, F., Schwope, A., \& Zavlin, V.~E.\ 2008, 40 Years of Pulsars: Millisecond Pulsars, Magnetars and More, 983, 354 

\bibitem[\protect\citeauthoryear{Neuh\"auser}{2001}]{neu2001} Neuh\"auser, R. 2001, AN, 322, 3

\bibitem[\protect\citeauthoryear{\"Ogelman et al.}{2005}]{ot2005} \"Ogelman, H., Tepedelenlioglu, E.  2005, astro-ph/0503215 

\bibitem[\protect\citeauthoryear{Pavlov et al.}{2004}]{pavlov2004} Pavlov, G.~G., Sanwal, D., \& Teter, M.~A.\ 2004, Young Neutron Stars and Their Environments, 218, 239 

\bibitem[\protect\citeauthoryear{Popov et al.}{2003}]{popov2003} Popov, S.~B., Colpi, M., Prokhorov, M.~E., Treves, A., \& Turolla, R.\ 2003, \aap, 406, 111 

\bibitem[\protect\citeauthoryear{Popov et al.}{2006}]{popov2006} Popov, S.~B., Turolla, 
R., \& Possenti, A.\ 2006, \mnras, 369, L23 

\bibitem[\protect\citeauthoryear{Posselt et al.}{2007}]{posselt2007} Posselt, B., Popov, S.B., Haberl, F., et al. 2007, Ap\&SS, 308, 171

\bibitem[\protect\citeauthoryear{Posselt et 
al.}{2008}]{posselt2008} Posselt, B., Popov, S.~B., Haberl, F., Tr{\"u}mper, J., Turolla, R., \& Neuh{\"a}user, R.\ 2008, \aap, 482, 617 


\bibitem[\protect\citeauthoryear{Poveda et al.}{1967}]{poveda1967} Poveda, A., Ruiz, J., \& Allen, C.\ 1967, Boletin de los Observatorios Tonantzintla y Tacubaya, 4, 86 

\bibitem[\protect\citeauthoryear{Reichen et al.}{1990}]{reichen1990} Reichen, M., Lanz, T., Golay, M., \& Huguenin, D.\ 1990, \apss, 163, 275 



\bibitem[\protect\citeauthoryear{Ruprecht et al.}{1981}]{ruprecht1981} Ruprecht, J., Balazs, B.~A., \& White, R.~E.\ 1981, Budapest: Akademiai Kiado, 1981,  

\bibitem[\protect\citeauthoryear{Schwope et al.}{2005}]{schwope2005}Schwope, A.D., Hambaryan, V., Haberl, F., Motch, C. 2005, A\&A 441, 597

\bibitem[\protect\citeauthoryear{Schwope et al.}{2007}]{schwope2007}Schwope, A.D.  Hambaryan, V., Haberl, F., Motch, C. 2007, Ap\&SS, 308, 619  

\bibitem[\protect\citeauthoryear{Stone}{1991}]{stone1991} Stone, R.~C.\ 1991, \aj, 102, 333 

\bibitem[\protect\citeauthoryear{Treves et al.}{2000}]{treves2000} Treves, A., Turolla, R., 
Zane, S., \& Colpi, M.\ 2000, \pasp, 112, 297 

\bibitem[\protect\citeauthoryear{van Kerkwijk \& Kaplan}{2007}]{vankerkwijk2007} van Kerkwijk, M.~H., \& Kaplan, D.~L.\ 2007, \apss, 308, 191 

\bibitem[\protect\citeauthoryear{van Kerkwijk 
\& Kaplan}{2008}]{vankerkwijk2008} van Kerkwijk, M.~H., \& Kaplan, D.~L.\ 2008, \apjl, 673, L163 

\bibitem[\protect\citeauthoryear{Yakovlev \& Pethick}{2004}]{yakovlev2004} Yakovlev, D.~G., \& Pethick, C.~J.\ 2004, \araa, 42, 169 

\bibitem[\protect\citeauthoryear{Walter}{2001}]{walter2001} Walter, F.~M.\ 2001, \apj, 549, 
433

\bibitem[\protect\citeauthoryear{Wang et al.}{2006}]{wang2006} Wang, C., Lai, D., 
\& Han, J.~L.\ 2006, \apj, 639, 1007 

\bibitem[\protect\citeauthoryear{Wielen}{1977}]{wielen1977} Wielen, R.\ 1977, \aap, 60, 263 

\bibitem[\protect\citeauthoryear{Woods \& Thompson}{2006}]{wood2006} Woods, P.~M., \& Thompson, C.\ 2006, Compact stellar X-ray sources, 547 

\bibitem[\protect\citeauthoryear{Zane et al.}{2005}]{zane2005} Zane, S., Cropper, M., 
Turolla, R., Zampieri, L., Chieregato, M., Drake, J.~J., 
\& Treves, A.\ 2005, \apj, 627, 397 

\bibitem[\protect\citeauthoryear{Zane et al.}{2006}]{zane2006} Zane, S., de Luca, A., Mignani, R.~P., \& Turolla, R.\ 2006, \aap, 457, 619 

\bibitem[\protect\citeauthoryear{Zane et al.}{2008}]{zane2008} Zane, S., Mignani, R.~P., 
Turolla, R., Treves, A., Haberl, F., Motch, C., Zampieri, L., 
\& Cropper, M.\ 2008, \apj, 682, 487 

\end{thebibliography}
\end{document}